\documentclass[journal=jacsat,manuscript=article]{achemso}
\usepackage{chemformula} 
\usepackage[T1]{fontenc} 
\usepackage{amsmath, amssymb, amscd, amsthm, amsfonts}
\usepackage{mathrsfs}
\usepackage{xcolor}
\usepackage{graphicx}
\usepackage{hyperref}
\usepackage{caption}
\usepackage{subcaption}
\usepackage{multicol}
\usepackage{algorithm}
\usepackage{algpseudocode}
\usepackage{tabularray}
\usepackage{array}
\usepackage{ulem}
\newcolumntype{P}[1]{>{\centering\arraybackslash}p{#1}}

\newcommand{\angstrom}{\mbox{\normalfont\AA}}

\author{Arsen Sultanov}
\affiliation[ICMPE]{Univ Paris Est Creteil, CNRS, ICMPE, UMR 7182, 2 rue Henri Dunant, 94320 Thiais, France}
\email{arsen.sultanov@cnrs.fr}
\author{Jean-Claude Crivello}
\affiliation[ICMPE]{Univ Paris Est Creteil, CNRS, ICMPE, UMR 7182, 2 rue Henri Dunant, 94320 Thiais, France}
\email{jean-claude.crivello@cnrs.fr}
\altaffiliation{Current address: CNRS-Saint-Gobain-NIMS, IRL 3629, Laboratory for Innovative Key Materials and Structures (LINK), National Institute for Materials Science (NIMS), 1-1 Namiki, 305-0044 Tsukuba, Japan}
\author{Tabea Rebafka}
\affiliation[LPSM]{
LPSM, Sorbonne Université, Université Paris Cité, CNRS, 75005 Paris, France
}
\altaffiliation{Université Paris-Saclay, INRAE, MaIAGE, Jouy-en-Josas, France}
\email{tabea.rebafka@sorbonne-universite.fr}
\author{Nataliya Sokolovska}
\email{nataliya.sokolovska@sorbonne-universite.fr}
\affiliation[LCQB]{LCQB, UMR 7238 CNRS, Sorbonne Université, 75005 Paris, France}

\title[Data-Driven Score-Based Models for Generating Stable Structures with Adaptive Crystal Cells]{Data-Driven Score-Based Models for Generating Stable Structures with Adaptive Crystal Cells }

\abbreviations{IR,NMR,UV}

\SectionNumbersOn 
\begin{document}

\begin{tocentry} 
\includegraphics[width=\textwidth]{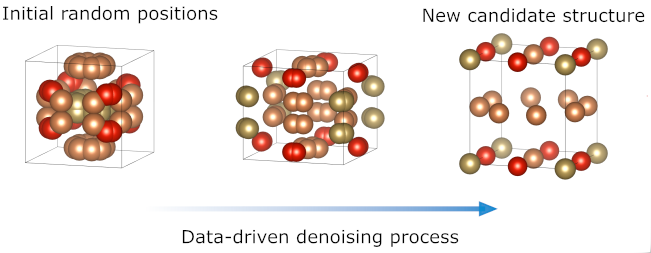} 
\end{tocentry}

\begin{abstract}
The discovery of new functional and stable materials is a big challenge due to its complexity. 
This work aims at the generation of new crystal structures with desired properties, such as chemical stability and specified chemical composition, by using machine learning generative models.  
Compared to the generation of molecules, crystal structures pose new difficulties arising from the periodic nature of the crystal and from the specific symmetry constraints related to the space group. 
In this work, score-based probabilistic models based on annealed Langevin dynamics, which have shown excellent performance in various applications,  are adapted to the task of crystal generation. 
The novelty of the presented approach resides in the fact that the lattice of the crystal cell is not fixed. During the training of the model, the lattice is learned 
from the available data, whereas during the sampling of a new chemical structure, two denoising processes are used in parallel to generate the lattice along the generation of the atomic positions. 
A multigraph crystal representation is introduced that respects symmetry constraints, yielding computational advantages and a better quality of the sampled structures.
We show that our model is capable of generating new candidate structures in any chosen chemical system and crystal group without any additional training.
To illustrate 
the functionality of the proposed method, 
a comparison of our model to other recent generative models, based on descriptor-based metrics, is provided.

\bigskip

\noindent\includegraphics[width=0.9\textwidth]{TOC.png}

 
\end{abstract}


\newpage
\section{Introduction}

The discovery of new materials with desired properties is challenging. We are particularly interested in the search for novel stable crystal structures. Discovering new stable crystal structures is of paramount significance, since there is an acute need to find new materials with certain characteristics. 
From the practical point of view, it is difficult to discover a new material with given properties. Traditionally, new materials are found through laborious experimental research, where human experts apply their knowledge, intuition, and understanding of mechanisms together with some simple heuristics to discover new chemical compounds. In recent decades with the advent of powerful computers and the density functional theory (DFT), it has become possible to estimate the stability of hypothetical crystal structures using high-throughput numerical simulations. This, in turn, allows new materials to be discovered through extensive brute-force screening. However, this approach is still far from being efficient, as the computation is extremely expensive in terms of CPU resources.

Machine learning offers a completely different approach. In essence, machine learning can  be thought of as automated learning from data. Methods can be designed to make predictions or decisions by learning structures and discovering patterns in existing data. More recent models are trained to generate new instances, with similar properties to those in the training data.  Applied in the context of materials discovery, generative machine learning methods can provide fast and viable solutions to problems that are otherwise expensive and difficult to solve.

\subsection{Background}

Given the huge search space in which new materials could be found, the application of machine learning methods can be of great help. 
However, this type of approach requires a large amount of data to explore the search space. In the case of the crystal structures, the available data is very limited. Another major problem is the representation of chemical crystal structures. Careful engineering of feature sets and efficient data representation are crucial for building scalable and effective machine-learning models.

Genetic algorithms are among the oldest statistical methods used in materials science. They have been considered for searching the complex chemical space, but their application is computationally too expensive for systems with more than two~different elements~\cite{Le16,Henson18}. Recent advances in machine learning methods propose surrogate algorithms for materials science applications that are cost effective compared to genetic algorithms. This research direction seems to be fruitful~\cite{Xie20}. Some promising results have been reported on the prediction of chemical and physical properties~\cite{Ward16}. The relationships between structure and properties have also been investigated~\cite{Dong19,Oliynyk17}. Note that there is no single optimal machine learning method for all materials science applications, so Support Vector Machines (SVM) were used to predict reaction outcomes of syntheses of templated vanadium selenites~\cite{Raccuglia16}. Random Forests (RF) were successfully applied to predict catalyst performance in C-N coupling reactions~\cite{Ahneman18}. 
Deep learning has been used to predict the reactivity of reaction combinations with high accuracy~\cite{Granda18}. A number of papers report promising results on machine learning for chemical synthesis~\cite{Moosavi19,Coley18,Voznyy19}. The XGBoost model was tested on a trial-error synthesis process of single crystals to predict their crystallisation propensity~\cite{Xie20}. Note, however, that it has also been reported that data mining methods for chemical applications, e.g. for the synthesis of inorganic-organic hybrid materials, are still rather limited~\cite{Raccuglia16,Moosavi19}. 

An important part of machine learning algorithms are generative models, which aim to produce new data points that resemble the data shown during training. Best known for their ability to generate realistic images, generative models are capable of mimicking many other types of data, including text, video and, most notably, chemical structures. Finding and extrapolating relevant patterns in data is at the heart of new data generation. Generative models have shown their efficiency and some promise for the generation of new materials. Most generative models are based on the construction of a material latent space to encode the information of the data set. This continuous material vector space is then used to generate new data points, such as crystal structures. It is also possible to introduce a mapping between a generated material and the desired property, leading to data generation with given properties~\cite{Kim20}. 

Recently, various generative deep learning models, have been adapted to the general task of suggesting new stable material candidates, most of which are based on either variational autoencoders (VAE)~\cite{imatgen, unet_original, unet, CDVAE} or generative adversarial networks (GAN)~\cite{crystalgan, cgan, cccgan}. Notably, a recent method explores the latent space of a trained VAE generative model aiming at targeted exploration of chosen chemical systems~\cite{composition_CDVAE}. 
Generative machine learning models are even more developed in the domain of molecular generation~\cite{Sousa2021, Kang2019}, with many examples relying on string representations of molecules~\cite{MolGPT, Sattarov2019}.
A number of deep learning architectures for materials science applications have been developed~\cite{Putin18,Lim18,Fuhr22,Pyzer22,Zuo21}. 
Some of the first attempts to generate new candidate structures using machine learning methods used 3D voxelised image representations~\cite{imatgen, unet_original, unet, cgan}, since the generative models were originally applied in the computer vision and image processing communities. Other approaches have used atomic coordinates and lattice parameters directly~\cite{crystalgan, cccgan}. However, an important problem is that these representations are not invariant to the isometries of the space.
Graph neural networks are widely and successfully exploited in molecular chemistry~\cite{shi, hoogeboom22} and also in crystal property prediction~\cite{cgcnn, dimenet++, gemnet, alignn} and do not suffer from this drawback. In addition, they can capture the local structure quite accurately. On the downside, they often struggle to capture the long-range dependencies that are crucial in periodic materials. 

Score-based models~\cite{DDPM,song,diffusion_survey} have been reported to achieve the state-of-the-art generative performance. These models are motivated by non-equilibrium thermodynamics and the main idea is to learn a process that gradually corrupts an input by adding Gaussian noise. The transformations used to reconstruct the input are learned. After a number of iterations, a score model is able to reconstruct the input from nothing but Gaussian noise. The score-based generative models have recently been developed for the generation of 3D molecules~\cite{shi,Xu22,hoogeboom22}. 

\subsection{Problem definition}

In this work the goal is to develop a model for the generation of new stable crystal structures that incorporates known symmetry constraints of chemical structures, while being more flexible and expressive than existing methods. Furthermore, we seek a method that generates structures with user-specified chemical composition (up to equivalent symmetric positions), so that it can be applied to any chemical system of interest. 
There are several major hurdles to the application of generative models in materials science. Firstly, the data available is very limited: the open-access Materials Project\cite{MP} database contains just over 33,~000 stable structures. This is much less than used for the training of generative models in other fields. In image processing, for example, there are databases containing several million observations or more. Secondly, the nature of the data is quite specific, as the structure should ideally be represented in a way that is invariant to the rigid transformations of space such as rotations and translations. Furthermore, an important characteristic of a crystal structure which makes it particularly difficult to generate but has to be taken into account, is its periodic nature.
 This challenge has recently been addressed by using graph representations of chemical compounds~\cite{Ren20,Baird22,Zhao21,Siriwardane22}. However, the proposed methods still lead to suboptimal performance. 


\subsection{Our contributions}

We challenge to overcome the drawbacks of the state-of-the-art methods, and our contribution is manifold:
\begin{itemize}
    \item First, we propose a score-based model for chemical structures with  periodic nature, namely crystals. To the best of our knowledge, we are the first to propose a method that applies the score-based model not only to the atomic site coordinates, but also to the lattice itself.  This double denoising process results in a better exploration of the search space and leads to a much more meaningful generation of candidate structures.
    \item Second, our model takes the number of symmetrically non-equivalent atoms per element as an input. Therefore it can be directly used for the task of targeted exploration of systems of any desired composition without additional training.
    \item  Third, our model respects the given constraints of a chosen space group, by using a novel symmetry-consistent multigraph description. 
    This helps to capture long-range patterns and improves the quality of the generated structure. 
    \item Finally,  the proposed approach is validated on the stable structures of the Materials Project database. Our numerical experiments consist of three parts. Firstly,
    the performance of our model is compared to that of alternative state-of-the art models using statistical metrics. Secondly, it is shown that our score-based model is able to generate existing crystal structures that were not seen by the algorithm during training. Lastly, examining  structures that are generated by our model  by computing their electronic and mechanical stability by DFT/phonon calculations, we identify original unknown structures of great relevance.
\end{itemize}

\subsection{Organization of the paper}

Section \ref{notations} introduces the crystal structure notations used in our paper. Section \ref{general_diffusion} provides a general formulation of the Score Matching with Langevin Dynamics (SMLD) model for new data generation.  In Section~\ref{our_model} we adapt this approach to the specific problem of generating new crystal structures.
Finally, Section~\ref{experiments} presents the results of numerical experiments carried out using our model.

\section{Notations}\label{notations}
A crystal structure is described by a unit cell (arrangement of atoms) and its Bravais lattice, which defines the periodic conditions.
The unit cell is a region of space bounded by a parallelepiped defined by three lattice vectors and the number, nature and position of the atoms contained in that region of space. A crystal structure is then generated by replicating and translating the unit cell according to its lattice vectors. 
Formally, we define a crystal structure as $M\,=\,(A, L, X, H) \in \mathbb{A}^N \times \mathbb{R}^{N \times 3} \times \mathbb{R}^{3 \times 3} \times \mathbb{H}$, where 
the following notations are used:

\begin{enumerate}
    \item $A\in \mathbb{A}^N$ denotes  the composition of a structure with $N$ atoms per unit cell. It indicates  the chemical element of every atom in the unit cell and $\mathbb A$ is the set of all   chemical elements.
    \item $L=(l_1, l_2, l_3) \in \mathbb{R}^{3\times 3}$ denotes the lattice matrix  defining the form of the unit cell.
The whole structure is invariant  to translations of the form $k_1 l_1 + k_2 l_2 + k_3 l_3$,  $(k_1, k_2, k_3) \in \mathbb{Z}^3$. Alternatively, the lattice can be defined by the lengths $a,b,c$ of its vectors and the angles $\alpha, \beta, \gamma$ between them,   denoted by $L_6 =(a,b,c,\alpha, \beta, \gamma)$. This representation is invariant to the choice of the coordinate system. When  angles are fixed,  we also use the representation given by $L_3=(\varphi(a), \varphi(b), \varphi(c))$,
    where $\varphi(x)=k\log(\exp(x/k)+1)$ for some positive real number $k$. Note that $L_3$ takes its values in $\mathbb R^3$, while the lengths $(a, b,c)$ are necessarily positive real numbers.
    \item $X \in \mathbb{R}^{N \times 3}$ is the matrix of atomic Cartesian coordinates of the atomic positions that are not equivalent with respect to the symmetry operations. Therefore, $XL^{-1}$ corresponds to the matrix of the atomic coordinates relative to the unit cell. Note that $X$ is usually much smaller than the full matrix of all atomic coordinates, since there are a lot of equivalent atomic positions in most structures.
    \item $H \in \mathbb{H}=\{H_1, ...,H_{230}\}$ denotes  the crystallographic space group defining  symmetry constraints for the structure \cite{bilbao}. A space group $H$ contains the symmetry operations $H = \{h_i, i \in 1,...,|H|\}$. 
    For a given space group, it holds that  $X$ and $h_i(X)$ define the same atomic positions, up to permutation, for all $i \in 1,...,|H|$.
    Moreover,  by  the symmetry constraints of  $H$, the position of one atom may determine the position of some of the other atoms in the unit cell. 
    Hence, storing all atomic positions is redundant, and in our model for every set of equivalent positions, we only keep track of the position of one of the atoms. 
    Some of the symmetry groups also put  constraints on the lattice. Depending on the space group $H$,   the lattice is bound to be either triclinic ($H_1, H_{2}$), monoclinic ($H_3, ...,H_{15}$),  orthorhombic ($H_{16}, ...,H_{74}$), tetragonal ($H_{75}, ...,H_{142}$), hexagonal ($H_{143}, ...,H_{195}$), or cubic ($H_{196}, ...,H_{230}$).
\end{enumerate}

\section{General score-matching with Langevin dynamics model}\label{general_diffusion}

Data is assumed to be a sample from an unknown probability distribution with density $p(x)$ defined over the data space. The goal is to learn how to sample from this distribution. The classical statistical approach is to construct a density estimate, say $\hat p$, and sample directly  from $\hat p$. Alternatively, one can use  Langevin dynamics of the form
\begin{equation}\label{langevin_sde}
    dX(t) = s\left( X(t) \right) dt + \sigma dB(t),
\end{equation}
where  $X$ is the process of interest, $t$ is the time, $B$ is a Wiener process and
 $s$ is the Stein score of the distribution $p$ defined by $s(x) = \nabla_x \log p(x)$. At  $x$, the score points in the direction of higher density. 
Intuitively, the process $X(t)$ tends to push towards the high density  regions of $p$.
More formally, one can show that, with $\sigma = \sqrt{2}$,  the invariant distribution of this process is $p$, i.e., as time $t$ becomes large,   the distribution of $X(t)$ tends to the distribution $p$. Thus, the process $X$ can be used  as a sampling method, since simulating the process $X(t)$ yields a distribution close to $p$, which corresponds to our goal.

From a physical point of view,   $-\log p(x)$ can be seen as an the energy function $U(x)$ of a system in  state $x$ and $p(x)$ is the corresponding Gibbs distribution for a given choice of the temperature. In this case, the score $s$ in the first term of the stochastic differential equation~\eqref{langevin_sde} corresponds to $-\nabla_x U(x)$, which pushes the system towards a minimum of its energy function, while the second term represents the external forces.
   
Our generative   model has two main parts:  training and  sampling. First, during the training, 
the score $s$ has to be learned from the available data set. Typically,  
a deep  neural network parameterized by $\theta$ is trained to give an estimate $s_\theta$. Second, new data points are sampled from $p$ using score matching and annealed Langevin dynamics.  An Euler scheme is used to obtain a numerical approximation of the Langevin process. 


\subsection{Score matching}

Let  $D = \{x_1, ..., x_n\}$ be a dataset of sample size $n$ with unknown distribution  $p$ and
  $p_0(\tilde x)=\frac1n\sum_{x \in D}\delta_{x}(\tilde x)$ the associated empirical distribution, where $\delta_{x}$ is the Kronecker delta. 
As $p_0$ is not differentiable, it cannot be used to construct an estimate of the score $s$. However, a smooth density estimate is given by the Gaussian kernel density estimator $\hat{p}_{\sigma}$ defined by
\begin{equation*}
    \hat{p}_{\sigma}(\Tilde{x}) = \dfrac{1}{n} \sum_{x \in D} f_{\mathcal{N}_{\mathrm{dim}(x)}(x, \sigma^2 I)}(\Tilde{x}),
\end{equation*}
where $f_{\mathcal{N}_{\mathrm{dim}(x)}(x, \sigma^2 I)}$ denotes the multivariate Gaussian density of dimension $\mathrm{dim}(x)$ 
with mean $x$ and covariance matrix $\sigma^2 I$, where $I$ is  the identity matrix and $\mathrm{dim}(x)$   the dimension of $x$.  
Denote $p_\sigma$ the limit of $\hat{p}_\sigma$ when $n \rightarrow \infty$, which can be expressed as
\begin{equation}
    p_{\sigma}(\Tilde{x}) = \int f_{\mathcal{N}_{\mathrm{dim}(x)}(x, \sigma^2 I)}(\Tilde{x})p(x)dx.
\end{equation}
Now, for training the score, say $s_{\theta}(\cdot, \sigma)$, which may be parameterized by a  neural network, the  straightforward objective function is  the mean squared loss between the score estimate and the score of $\hat{p}_\sigma$ given by
\begin{equation}
    \mathbb{E}_{\Tilde{x} \sim \hat{p}_\sigma}\left[\left\lVert s_{\theta}(\Tilde{x}, \sigma) - \nabla_{\Tilde{x}} \log \hat{p}_\sigma(\Tilde{x}) \right\rVert_2^2\right].
    \label{eq:expect1}
\end{equation}
However, the calculation of  $\nabla_{\Tilde{x}} \log \hat{p}_\sigma(\Tilde{x})$ is expensive, but instead, one can use another simpler objective given by
\begin{equation}\label{eq:expect2}
    \ell_{\sigma}(\theta) = \mathbb{E}_{x\sim p_0, \Tilde{x}|x\sim \mathcal{N}_{\mathrm{dim}(x)}(x, \sigma^2 I)}\left[\left\lVert s_{\theta}(\Tilde{x}, \sigma) - \nabla_{\Tilde{x}} \log f_{\mathcal{N}_{\mathrm{dim}(x)}(x, \sigma^2 I)}(\Tilde{x}) \right\rVert_2^2\right].
\end{equation}
It has been shown that these two objectives, equations~\ref{eq:expect1} and \ref{eq:expect2}, are equivalent\cite{score_proof}, where  two objectives $\ell_1$ and $\ell_2$ are said to be equivalent if $\ell_1 = \alpha \ell_2 + \beta$ for some $\alpha > 0$, $\beta \in \mathbb{R}$. In particular that means that gradient descent optimization will yield the same result when performed on either objective.

The gradient on the right-hand side of \eqref{eq:expect2} has a simple analytical expression, which is convenient for its numerical evaluation. 
However, the expectation is not explicit, but can be approximated by Monte-Carlo simulations. 
In our case, we use minibatch sampling to estimate $\ell_{\sigma}(\theta)$. Let $\{x_1, ..., x_m\}$ be a  minibatch of size $m$, then for every $x_i, i \in 1,...,m$ we sample one  data point $\Tilde{x_i} \sim \mathcal{N}_{\mathrm{dim}(x_i)}(x_i, \sigma^2 I)$ that is used to compute an empirical counterpart of $\ell_{\sigma}(\theta)$. That is, 
the objective becomes
\begin{equation}\label{loss_term}
    \hat{\ell}_{\sigma}(\theta) = \dfrac{1}{m}\sum_{i=1}^m\left[\left\lVert s_{\theta}(\Tilde{x}_i, \sigma) - \nabla_{\Tilde{x}_i} \log f_{\mathcal{N}_{\mathrm{dim}(x_i)}(x_i, \sigma^2 I)}(\Tilde{x}_i) \right\rVert_2^2\right].
\end{equation}

When the sample size $n$ tends to infinity, the minimizer $s_{\theta^*}(\Tilde{x}, \sigma)$ of~\eqref{loss_term} equals the score of the distribution $p_{\sigma}$ almost surely.


Clearly, the choice of the value of $\sigma$ has an impact on the result. If $\sigma$ is small, $\hat{p}_\sigma$ (and $p_\sigma$) is a multimodal density, where all mass is concentrated on small neighbourhoods of the data points. As a consequence,   an iterative sampling procedure is likely to get stuck in a mode with no means to escape from it. On the contrary, a large value of $\sigma$ implies a flat density $\hat{p}_\sigma$, which favours the exploration of the whole data space during sampling, but which is a rather biased estimate of the target distribution $p$. 

To address this issue, one may use Langevin dynamics \cite{song}, that consists in combining the solutions for a predefined set of decreasing noise levels $\sigma_1 > ... > \sigma_F > 0$ with $F$ being the number of noise levels. This approach provides   stable score estimates, and during sampling   all modes of the distribution may be reached.  
The score $s_{\theta}$  is trained by minimizing  the following objective, which is a weighted average of the loss functions defined above, given by
\begin{equation}\label{total_loss}
    \mathcal{L}(\theta) = \sum_{k=1}^F w_k \hat{\ell}_{\sigma_k}(\theta).
\end{equation}
By choosing the weights as $w_k = \sigma_k^2$, the terms in the sum in~\eqref{total_loss} are of nearly the same order of magnitude \cite{song}. As such, no particular noise level is prioritized over others. For computational reasons, we use a unique neural network $s_\theta$ to estimate the score for all noise levels. To distinguish the scores corresponding to different noise levels, we pass the noise level $\sigma_k$ as one of the inputs to the neural network that provides the estimate of the score of $p_{\sigma_k}$.

\subsection{Sampling}

The trained estimator $s_\theta$ is  used for sampling. The Langevin dynamics process in~\eqref{langevin_sde}  can be approximated by the Euler scheme  
\begin{equation}\label{euler}
     x_{t} = x_{t-1} + \frac{\epsilon}{2}s(x_{t-1}) + \sqrt{\epsilon}z_t,
\end{equation}
where $z_t$ has standard normal distribution $\mathcal{N}_{\mathrm{dim}(x_t)}(0, I)$ and $\epsilon$ is the step size. Under convenient assumptions, the scheme converges to the stationary distribution of the Langevin dynamics Stochastic Differential Equation (SDE). Using this scheme and starting from  some randomly initialized data point $x_{1,0}$, 
  $T$ iterations according to equation~\eqref{euler} are performed. The final value $x_{1,T}$ can be considered as a realization of $p_{\sigma_1}$ and is used as the initial value for the sampling at the next noise level, that is, we set $x_{2,0}=x_{1,T}$. The noise parameter $\sigma_1$ should be large enough, so that the process is unlikely to get stuck in a local maximum of the $p_{\sigma_1}$, and the data space can be explored rapidly. The next Langevin simulation samples from $p_{\sigma_2}$ and the result is used as the initialization for the sampling from $p_{\sigma_3}$ and so on, until we obtain a sample from $p_{\sigma_F}$. Since we need the final distribution $p_{\sigma_F}$ to be close to $p$, one should choose $\sigma_F$ to be small enough.


\section{New crystal score-based model}\label{our_model}

Here we develop a score-based model specifically tailored for the generation of crystal formations. Our approach is such that  the user can choose the elements and their composition in a new structure that is generated by the model. This offers the possibility of a targeted search taking into account  prior expert knowledge or the practitioner's intuition. More precisely, to sample a new structure $M=\,(A, L, X, H)$ the user first chooses the composition $A$ and also the symmetry space group $H$, which helps the algorithm to sample realistic structures.
 
We restrict our attention to  space groups whose lattices are neither  triclinic or monoclinic.
For all these crystal systems, i.e., for $H\in\{H_{16},...,H_{230}\}$, the angles between the lattice vectors are fixed in the conventional lattice representation. Thus,  
it is sufficient to consider the generation of the lengths of the lattice vectors. As mentioned above, the transformed vector lengths $L_3$ are real-valued, which is convenient for adding Gaussian noise to the denoising process.

\begin{figure}
     \centering
     \includegraphics[width=.8\textwidth]{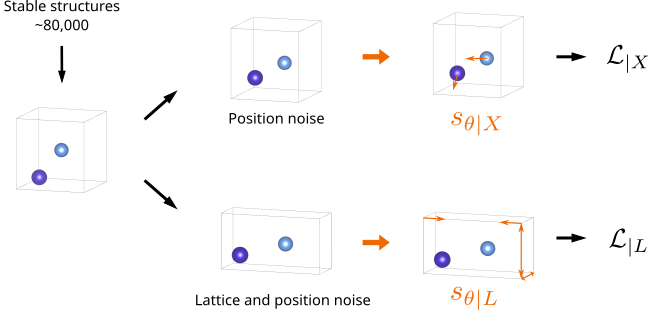}
     \caption{Model training. While learning the positions score $s_{\theta|X}$, the lattice remains unchanged. While learning the lattice score $s_{\theta|L}$, both the lattice and the atomic positions are perturbed. Orange arrows correspond to the score estimates given by the model as shown on Figure~\ref{score}.}
     \label{training}
\end{figure}
 
Now, in our model, the Langevin dynamics process is applied to both the atomic coordinates $X$ and the lattice parameters $L_3$, as shown in Figure~\ref{training}.
Note that $X$ and $L_3$ are objects of different nature which is new compared to existing score-based models, where the denoising process typically acts only on a single type of object. To handle this issue, we consider the processes on  $X$ and $L_3$ separately. That is, the score $s$  is separated into two parts,   one related to the atomic positions, and another corresponding to the lattice. We note them $s_{|X}$ and $s_{|L_3}$ respectively. Likewise, the global  loss function $\mathcal L$ introduced in \eqref{loss_term} can be split into two terms, say $\mathcal{L}_{|L}$ and $\mathcal{L}_{|X}$, such that the two scores  $s_{|X}$ and  $s_{|L_3}$ can be learned separately. More precisely, when training the score estimate $s_{\theta|X}$,   noise is added only to the Cartesian atomic coordinates, while the lattice remains unchanged. 
Fixing the lattice here makes sense from a physical standpoint, because otherwise  the objective would depend on the choice of the lattice.
However, for  the training of  $s_{\theta|L_3}$,   noise can be  added  to both the atomic coordinates and the lattice parameters, since the deformation would depend on the lattice anyway.

To obtain a flexible model, we choose different step sizes and noise levels for each of the Langevin dynamics processes in our model. Let  $(\epsilon_{X,k})_k$ and $(\epsilon_{L,k})_k$ be the step size sequences for the atomic coordinates and the lattice parameters, respectively. They are defined as
\begin{align*}
    \epsilon_{X,k} &=\gamma_{X}\; \sigma_{X, k}^2 / \sigma_{X, F}^2, \qquad
    \epsilon_{L,k} = \gamma_{L}\; \sigma_{L, k}^2 / \sigma_{L, F}^2,
\end{align*}
where $\{\sigma_{X,i}\}_{i=1}^{F},\gamma_X$ and $\{\sigma_{L, i}\}_{i=1}^{F},\gamma_{L}$ are two different sets of parameters.
For the sake of faster convergence we choose low temperatures for the lattice Langevin process by additionally multiplying the noise term by a factor of $0.15$. The resulting lattices are likely to be rather close to a local maximum of the associated probability density function. For the atomic positions, we keep the normal temperature for a better exploration of the sample space. 

\begin{figure}
     \centering
     \includegraphics[width=\textwidth]{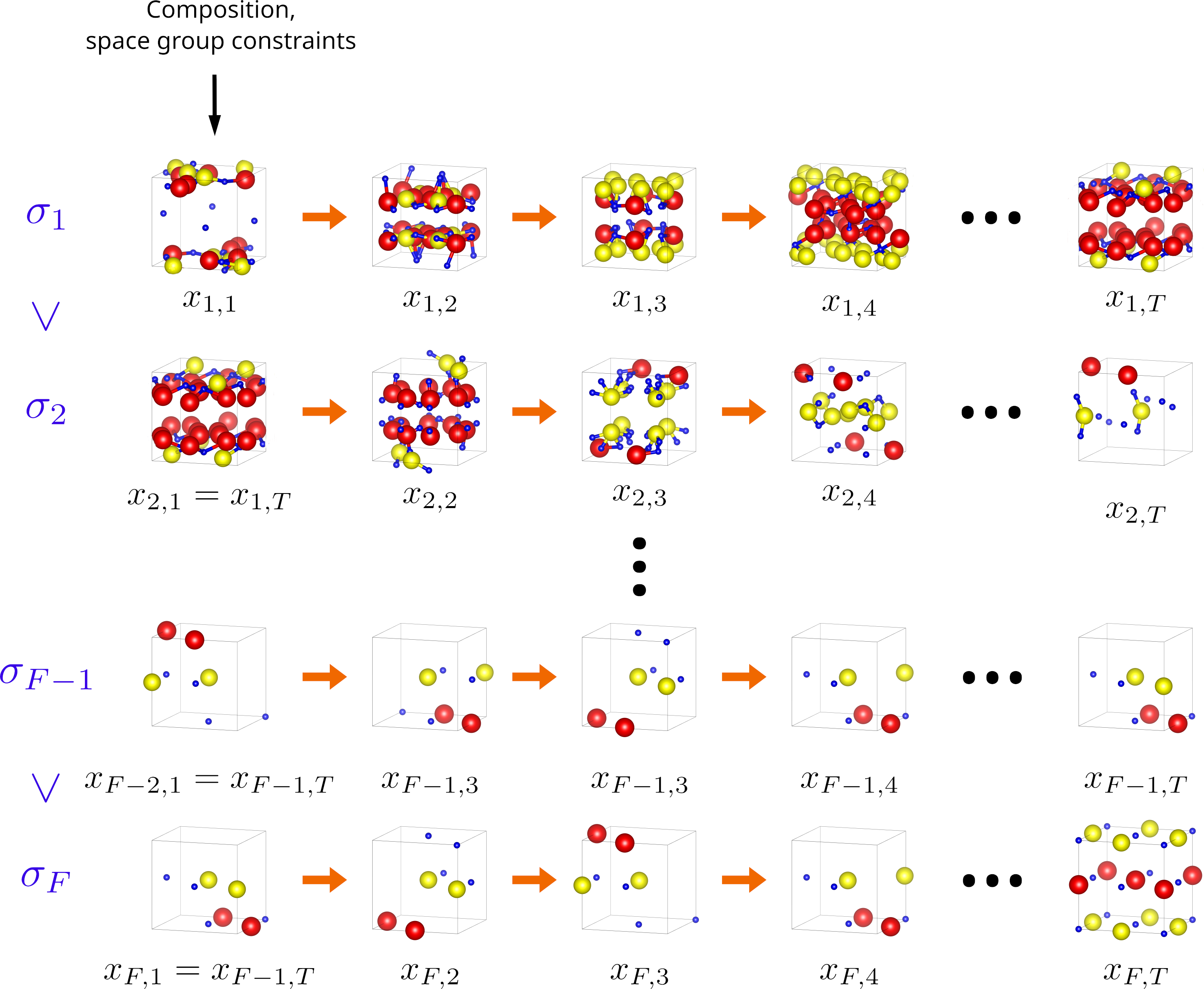}
     \caption{Sampling with annealed Langevin dynamics. Orange arrows correspond to the steps of the Langevin dynamics simulations, as shown in Figure~\ref{score}. The sample from the most perturbed distribution, corresponding to $\sigma_1$ is used as initialization for the subsequent dynamics, denoted as $\sigma_2$.}
     \label{sampling}
\end{figure}

As illustrated in Figure~\ref{sampling}, the sampling process  starts off with a randomly initialized structure, with lattice parameters drawn from a Gaussian distribution and 
atomic coordinates sampled  uniformly in the unit cell.
 Then, one iteration of the equation \eqref{euler} is done in two steps. First, the atomic positions are updated, and then a step over the lattice parameters $L_3$ is done. Our learning procedure is drafted as Algorithm~\ref{alg:cap}. 
 
Note that the actual number of  atoms in the structure is not fixed due to the symmetry constraints of the chosen  space group $H$. 
 For example, if an atomic position $x$ coincides with all of its symmetric positions, i.e., $h(x)=x$ for $h\in H$, then there is a single atomic site corresponding to $x$ in $A$. However, if  all  symmetric positions are different, that is, $h(x) \neq \tilde h(x)$ for all $h,\tilde h  \in H$ and $h\neq\tilde h$, then there are $|H|$ atomic sites in the structure corresponding to $x$. Therefore, as the atomic positions are modified during sampling, it may happen that the number of atoms changes.

\begin{algorithm}\label{algo}
\caption{Sampling with Langevin dynamics}\label{alg:cap}
\begin{algorithmic}
\Require $\{\sigma_{X, k}\}_{k=1}^{F},\gamma_X$ \Comment{Position denoising process coefficients} 

\hspace{1.47em} $\{\sigma_{L, k}\}_{k=1}^{F},\gamma_{L}$ \Comment{Lattice denoising process coefficients}

\hspace{1.47em} $T$ \Comment{Step count per noise level}

\smallskip

\hspace {-1.4cm} \textbf{Output:} $X, L$ \Comment{Atomic positions, lattice parameters}

\smallskip
\State Initialize ${X}_{1,0}$, ${L}_{1,0}$ \Comment{Initial atomic coordinates and lattice parameters}
\For{$k \gets 1$ to $F$}
  \State $\epsilon_{X,k} \gets \gamma_{X}\; \sigma_{X, k}^2 / \sigma_{X, F}^2$ \Comment{Atomic positions step}
  \State $\epsilon_{L,k} \gets \gamma_{L}\; \sigma_{L, k}^2 / \sigma_{L, F}^2$ \Comment{Lattice parameters step}
  \For{$t \gets 1$ to $T$}
      \State Draw $z_{X,t} \sim \mathcal{N}_{\mathrm{dim}(X)}(0,I)$
      \State Draw $z_{L_3,t} \sim \mathcal{N}_3(0,I)$
      \State ${X}_{k,t} \gets {X}_{k,t-1} + \frac{1}{2}\epsilon_{X,k} s_{\theta|X}({X}_{k,t-1}, {L}_{t-1}, \sigma_{X,k}) + \sqrt{\epsilon_{X,k}}z_t$
      \State ${L}_{k,t} \gets {L}_{k,t-1} + \frac{1}{2}\epsilon_{L,k} s_{\theta|L}({x}_{k,t-1}, {L}_{t-1}, \sigma_{L,k}) + 0.15\sqrt{\epsilon_{L,k}}z_{l,t}$
  \EndFor
  \State ${X}_{k+1, 0} \gets {X}_{k, T}$
  \State ${L}_{k+1, 0} \gets {L}_{k, T}$
\EndFor
\end{algorithmic}
\end{algorithm}

\subsection{Multi-graph representation}
The data points $(A,L,X,H)$ are complex objects that are not conventional for machine learning. In addition, using the atomic coordinates $X$ is prohibitive because this representation is not invariant to rotations and translations. Thus, the choice of an appropriate data representation is an important issue.

First, we denote by $X'$  the coordinates of all positions obtained by applying symmetry operations $h \in H$ to the atomic positions in $X$, which contain only one position for each set of equivalent positions in the real structure. Note that the number of coordinates in $X'$ can be larger than the number of actual atoms in a structure, since some of the positions in $X'$ may correspond to coinciding symmetric positions.

As in some other material generation models \cite{shi, CDVAE}, we adopt a multi-graph crystal representation $(V,E,W)$, which has the desired invariance properties, while preserving all structural information. In our graph representation, however, the graph vertices $V$ correspond to atomic positions in $X$, i.e., non-equivalent atomic positions within the unit cell.
The edges $E$ correspond to the interatomic bonds, such that one end of the bond is in $X$ and the other is in $X'$. However, since there is a single node for each set of equivalent positions in $X'$, these bonds create multiple edges and self-loops in the graph representation. Another source of multiple edges and self-loops is the periodicity of the structure since each atom can be connected to multiple copies of each atom obtained by the periodicity.
$W$ are the angles between these edges, i.e., angles $W(i,j,k)=\alpha_{ijk}$ formed by triplets of nodes $(i,j,k)$ such that $(i,j)\in E$ and $(j,k) \in E$. 

Introducing symmetry relations into the multigraph helps to account for the long-range atomic relationships within the structure.

Additional details can be found in the Supplementary Information.




\subsection{Graph neural network}
\begin{figure}
     \centering
     \includegraphics[width=\textwidth]{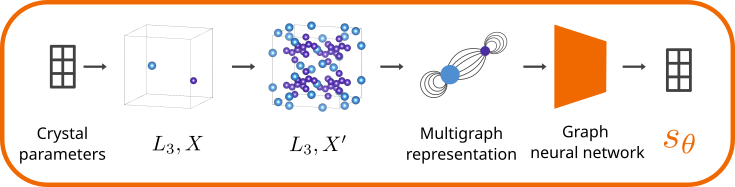}
     \caption{Score estimation, representing the orange arrow in Figures~\ref{training} and \ref{sampling}. The crystal structure is defined by its lattice parameters $L_3$ and the atomic positions $X$. All symmetric positions $X'$ are added. A multigraph is then constructed and passed to a graph neural network estimator that outputs the score.}
     \label{score}
\end{figure}

With the multigraph representation at hand, it is natural to use Graph Neural Networks (GNNs), a class of deep neural networks that operate on graph data.
The forward pass of a GNN generally consists of iteratively updating the multigraph representation by passing messages from each node to its neighbours.
In the context of crystal generative models, a GNN learns an appropriate edge output to estimate the score function, as elaborated further below.  
All the steps of the score evaluation are illustrated on Figure~\ref{score} and are represented by the orange arrow in Figures~\ref{training} and \ref{sampling}.

Here, we describe the estimation of the score $s_{|X}$, while the case of $s_{|L}$ is presented in Section~\ref{subsec:latticediffusion}.
Recall that for a given pair of vertices, say $i$ and $j$, there may be multiple edges $e_{ijk}$ due to the periodicity of the structure.  For each edge $e_{ijk}$, denote by $dx(e_{ijk})\in\mathbb R^3$  the vector of  the differences between the coordinates of the associated atomic sites. 
Let $w_{ijk}$ be the real output provided by the GNN for edge $e_{ijk}$. Then, the score of the atomic coordinates of a given node $i$ with coordinates $x_i$ is obtained as a weighted sum of normalized vectors $dx(e_{ijk})$ over all edges  originating at this node, that is, 
\begin{equation*}
     s_{\theta|X}(x_i) = \sum_{j \in N(i)}\sum_{k: e_{ijk} \in E(i,j)}w_{ijk}\frac{dx(e_{ijk})}{\lVert dx(e_{ijk})\rVert},
\end{equation*}
where $N(i)$ is the set of neighbours of node $i$ and $E(i,j)$ is the set of multiple edges between  $i$ and $j$.
Since we want a unique score vector for each set of symmetry equivalent atomic positions,   the scores for  nodes $x_i$ that were added by symmetries are not evaluated. In other words, the score is calculated only for the positions in $X'$ that are also present in~$X$.

A final post-sampling step is added to the procedure, which removes some of the atoms in $X'$ that we consider to come from symmetry operations applied to special positions in the unit cell. More precisely, any two atoms obtained by applying symmetry operations to an atomic site are considered to be the same atom if the distance between them is less than a given threshold, which we fix to $0.3\,\angstrom$. Thus, it is the number of non-equivalent atomic positions that is fixed, not the exact composition.

 Our implementation uses a Dimenet++ graph neural network \cite{dimenet++} as the estimator for our model, which
 uses Bessel basis representation for the interatomic distances and Fourrier-Bessel basis representation for the angles between bonds in its message passing function. The original Dimenet++  aggregates the messages in the atoms and performs pooling over the node features at the end, producing an output in the form of a vector for the whole structure. 
 However, our method requires an output for each edge in the graph, so we modified the output blocks by replacing the dense layers on the atomic features with dense layers on the edge features and removing the pooling.

\subsection{Crystal cell generation}\label{subsec:latticediffusion}

The lattice, which defines the periodicity of the crystal structure, is not unique. Usually,  either the conventional unit cell or the primitive unit cell is used. The former can have up to 4  
times more atoms in face-centered cubic unit cell, but the constraints it has to respect are much simpler than those of the primitive cell. There is an invertible mapping between the two types of the unit cell, so we assume that for each primitive unit cell there is a corresponding conventional unit cell and vice versa.

As the symmetry constraints are easier to handle in  the conventional unit cell, we consider a Langevin process acting on the lattice parameters of the conventional unit cell and aim to learn to denoise them. However, a large number of atoms in the unit cell  implies more stringent requirements in terms of memory and the execution time for the GNN. Therefore, to estimate the score of the perturbed conventional unit cell parameters, we still use  the multigraph representation associated with the corresponding primitive cell which we feed to the network. 

During   training, the noise $\Delta L_3$, which is added to the lattice representation $L_3$, has to be computed. 
The aim of lattice score training  is such that the GNN   computes an estimate  of $s_{|L_3} = \dfrac{\Delta L_3}{\sigma_k}$. 
Note that changing the lattice   also changes the Cartesian coordinates of the atomic sites, and   the estimate of $\Delta L_3$ is based on the  changes in relative distances between atoms induced by the deformation of the lattice.

More specifically, given a multigraph representation $(V, E, A)$ of the structure, for each edge $e \in E$ the network provides the amount of contraction/dilatation $\Delta d_{e}$ of the distance $d_e$ between the atoms associated with edge $e$ due to the lattice deformation. Independently, the gradient $G_e= \nabla_{L_3}d_e$ of $d_e$ with respect to the conventional lattice parameters $L_3$ is  approximated numerically. Indeed, for a sufficiently small lattice noise $\Delta L_3$, it holds that 
    $\langle G_e, \Delta L_3\rangle \approx \Delta d_e,$
where $\langle a,b \rangle$ denotes the scalar product between $a$ and $b$.
So, $\Delta L_3$ is naturally approximated by $\dfrac{\Delta d_{e,\theta}\, G_e}{\lVert G_e \rVert^2_2}$, a colinear vector with the gradient that changes the edge length by the predicted amount. We then take the mean of these estimates for all edges $e$, yielding the final estimate of the noise of the lattice parameter given by
\begin{equation*}
    \Delta L_{3,\theta} = \frac{1}{|E|} \sum_{e \in E} \dfrac{d_{e,\theta}\, G_e}{\lVert G_e \rVert^2_2}.
\end{equation*}
The lattice score estimate $s_{\theta|L}$ is then given by $s_{\theta|L} = \dfrac{\Delta L_{3,\theta}}{\sigma_k}$.

\section{Numerical results}\label{experiments}
In this section the performance of our model is assessed. First, it is compared to  other recently developed generative models via structure descriptor-based metrics on a set of structures sampled by the different models. Then, the model's capacity of generating known and unknown structures is investigated for a variety of chemical systems.

\subsection{Materials project data}\label{mp_data}
Our study uses most of the nearly stable structures (energy above hull is below $0.1 eV/$atom) available in the Materials Project~\cite{MP} (we accessed the data in February 2022). There are $86,508$ structures in total in our experiments.

Each structure is stored in a CIF file containing a matrix $L$ defining the lattice, and a matrix $X$ defining the atomic positions and the composition $A$. The space groups $H$ were identified using the Phonopy package \cite{phonopy}. It is important to emphasise that we did not use any additional information about the chemical properties of the structures other than the atomic number to identify the element species.

\begin{figure}
     \centering
     \includegraphics[width=.75\textwidth]{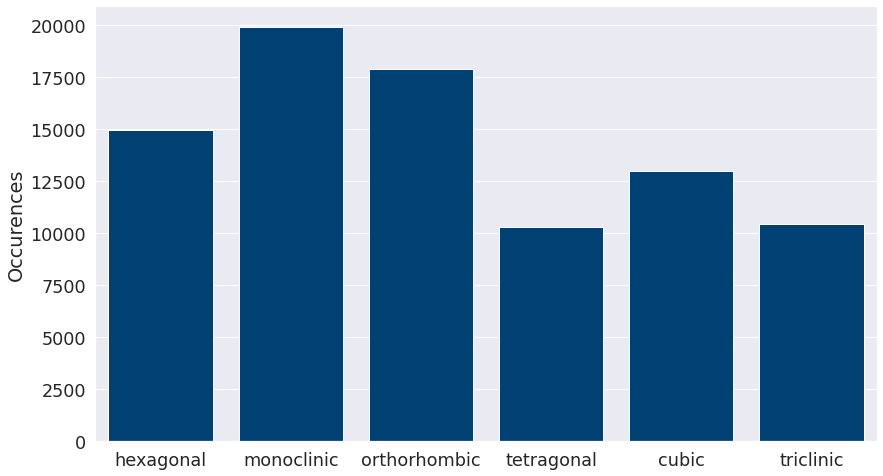}
     \caption{Distribution of crystal families in the dataset used in training. Monoclinic and triclinic systems are excluded from the dataset when training $s_{\theta|L_3}$.}
     \label{systems}
\end{figure}

\begin{figure}
     \centering
     \includegraphics[width=.75\textwidth]{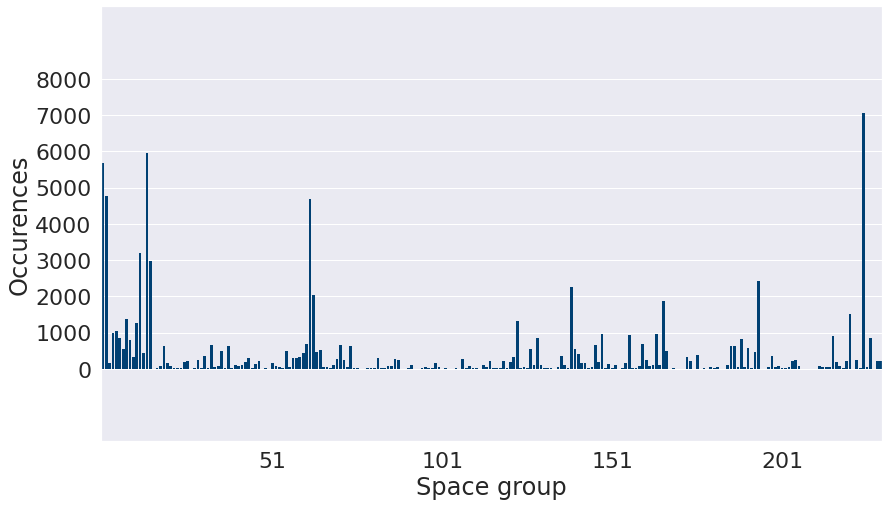}
     \caption{Distribution of the space groups of the structures in the dataset. }
     \label{space_groups}
\end{figure}

\begin{figure}
     \centering
     \includegraphics[width=.75\textwidth]{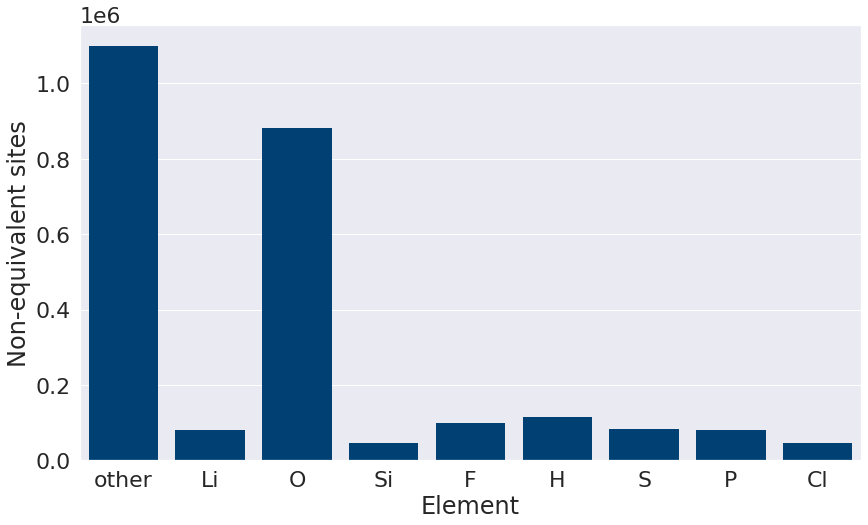}
     \caption{Number of non-equivalent atomic sites by element in the data set.}
     \label{elements}
\end{figure}

As shown in Figure~\ref{systems}, $30,358$ structures belong to the monoclinic and triclinic crystal families.
While the atomic position score $s_{\theta|X}$ is learned on the entire dataset, 
 the lattice score $s_{\theta|L_3}$ does not use  monoclinic and triclinic crystals in training, as we do not consider crystal families with variable angles in conventional lattice representation.  
Furthermore, the distribution of the space groups of the structures, displayed  in Figure~\ref{space_groups},
is far from uniform, with a considerable number of space groups not being represented in the data at all.
We also see that the dataset is heavily biased towards the oxygen element, which occupies $35\%$ of the atomic sites and with $40,282$ structures containing this atom (Figure~\ref{elements}).

\begin{figure}
     \centering
     \includegraphics[width=.75\textwidth]{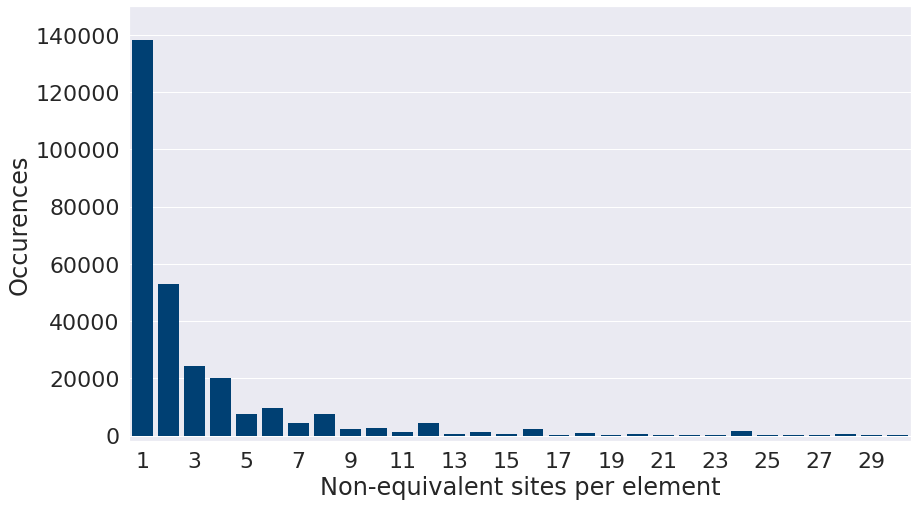}
     \caption{Distribution of the number of non-equivalent atomic sites per element in the data set.}
     \label{mult}
\end{figure}

As our model allows the user to specify the desired number of non-equivalent atomic sites, we examine the distribution of the number of non-equivalent atomic sites per element in the database. As shown in Figure~\ref{mult}, in most cases there are very few non-equivalent sites per element in any structure, mostly  only one or two  per element.
Based on this observation, it is reasonable to expect that new stable compounds will also be of rather simple nature. 

\subsection{Comparison with other models}
For a comparative study, we consider two  recent state-of-the-art crystal generative models as a baseline. First, the Crystal Diffusion Variational Autoencoder (CDVAE) \cite{CDVAE} is a variational autoencoder  that uses a diffusion process to denoise atomic coordinates from a latent representation. In that respect, this model is the most similar to ours. The second one is the Physics Guided Crystal Generative Model (PGCGM) \cite{PGCGM}, which uses a Generative Adversarial Network  to sample crystal structures  within a given set of 20 space groups.

For a fair comparison, all  models  should  be trained on the same dataset. We trained CDVAE and our model on the so-called MP-20 dataset proposed in the CDVAE paper \cite{CDVAE}, which consists of the Materials Project structures that have at most 20 atomic sites in its unit cell. We also removed the structures belonging to monoclinic and triclinic space groups, thus reducing the number of materials in the database from $45,229$ to $36,334$. We used the same 60:20:20 split of the data into training, test and validations sets to train the CDVAE and our model. The PGCGM material generator is publicly available albeit lacking the code to train on new data. It is known that PGCGM is trained on data collected from Material Project, Inorganic Crystal Structure Database (ICSD)\cite{ICSD} and Open Quantum Materials Database (OQMD)~\cite{OQMD}.
 


Concerning sampling, a major difference between our model and the other two is that  our model takes the chemical composition (up to the number of symmetrically equivalent atomic sites) and the space group operations as input.
Therefore, when sampling with our model, we feed the model with the composition and space group of one of the materials in the test set.

For each model, 1000 structures are  generated 
and we compute the corresponding coverage-recall (COV-R) and coverage-precision (COV-P) metrics proposed in the CDVAE paper \cite{CDVAE}, that are based on the comparison of the CrystalNN and Magpie fingerprints of the generated structures and the structures in the test set. The larger  the metrics, the better generated structures imitate the test data. More precisely, COV-P is defined as the proportion of generated structures that have close neighbours in the feature space among the test structures. Inversely, COV-R is the proportion of test data structures that are close  to the generated structures. Notably, COV-P heavily depends on the  size of the test dataset, while  COV-R heavily depends on the number of sampled structures. 

Inspired by the standard for the image generation task, Fr\'echet inception distance  metric \cite{TTUR}, we propose to consider two similar metrics for  crystal structures. The metric assumes that the structure descriptors are normally  distributed for both the generated and the test data and computes the Fr\'echet distance between those distributions. The lower the distance, the more similar the test and generated data are. We consider CrystalNN as the first descriptor, and the activations of the second to last layer of the Crystal Graph Convolutional Neural Network (CGCNN)~\cite{cgcnn} as the second. The distributions were calculated based on 1000 random samples for each model.

The results are displayed in the Table~\ref{tab:comparison}, where we note these metrics FD for Fr\'echet Distance. In case of the PGCGM model, since the model generates structures from only 20 space groups, we also compute the comparison metrics with the test data containing only structures from the same 20 space groups, the result shown in numbers between parentheses. The sharp drop in COV-P can be explained by the dependence of the metric on the size of the test data set.
We see that our model performs well across the board outperforming or nearly matching the performance of the other models.
 
\begin{table}[htbp]
    \begin{tblr}{
      colspec = {X[c,t]X[c]X[c]X[c]X[c]},
      stretch = 0,
      hlines = {1.pt},
      vlines = {1.pt},
    }
        Model   & COV-P & COV-R & FD CrystalNN & FD CGCNN \\\\
        This work & 0.972 & 0.965 & 0.102 & 0.682 \\
        PGCGM & 0.944 (0.864) & 0.76 (0.765) & 0.225 (0.303) & 4.359 (4.209) \\  
        CDVAE & 0.982 & 0.862 & 0.199 & 0.703 \\  
        1000 stable structures & 0.998 & 0.968 & 0.012 & 0.044 \\
    \end{tblr}
    \caption{Comparative study of our model to PGCGM and CDVAE using various metrics.  Values of the metrics for 1000 randomly picked stable structures.}
    \label{tab:comparison}
\end{table}

Finally, to examine the validity of the proposed metrics, we also evaluate these metrics for 1000  stable structures randomly chosen from the training dataset. Since we can assume that these structures come from the same distribution as the test structures, the corresponding structure descriptors should also have close distributions, resulting in close to perfect metric values. We observe from Table~\ref{tab:comparison} that the results are matching these expectations, supporting the validity of the chosen metrics.

\subsection{Sampling known stable structures}
The goal here is to assess the ability of our model to generate stable structures, which are not used during training. 
Here and further below we used our model trained on the complete Materials Project dataset as described in the Section \ref{mp_data}. First we did a 9:1 train-test split, using the test data to track the loss and choose the stopping point.

Some of the experimentally observed structures are not included in the training data set, in particular those structures that are unstable at $0 K$ and no pressure. Taking advantage of this fact, we attempt to sample these structures with our model by providing the correct space group and elemental composition, that is, the number of non-equivalent positions for each element. 
This way, we hand-picked 36 structures that have been observed experimentally in the Materials Project, but are not included in the training set. 
These 36 structures have been chosen to provide a set of samples as representative as possible of the whole chemical space (different crystal families and different types of chemical bonds). 
For each structure, we then generate 50 samples with the same number of non-equivalent positions, with the space group symmetries fixed in advance, and perform DFT relaxation on the output structures. This results in several original structures being found in the generated samples, some of them are shown in the Table~\ref{tab:reconstruction}. The complete list of tested structures with their material project IDs can be found in the Supplementary Material.

In total, we were able to reconstruct 11 of the 36 structures, while others are close but have different atomic arrangements. Concerning the lattice parameters in the reconstructed structures they may be slightly different from those in the original structures, but these deviations can simply be corrected by applying the DFT relaxation scheme.

Thus, we conclude that our model is able to reproduce known structures with a large number of different space groups across all targeted crystal families. 
In particular, the lattice denoising is indeed able to capture the desired lattice shape, while adapting to the constraints imposed by the chosen space group.

\begin{table}[htbp]
    \begin{tblr}{
      colspec = {X[c,t]X[c]X[c]X[c]X[c]X[c]},
      stretch = 0,
      hlines = {1.5pt},
      vlines = {1.5pt},
    }
        Elements & Crystal family and space group number& Generated structure before DFT relaxation & Lattice parameters of the sampled structure, $\angstrom$ & Lattice parameters according to Materials project, $\angstrom$ \\
        Li-Mg-P & Cubic, 216 & {\includegraphics[width=1.5cm]{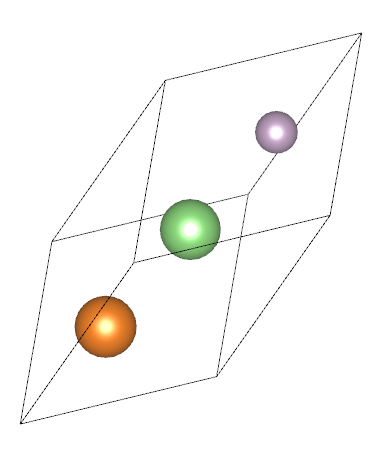}}
            & 4.43 & 4.08 \\
        Ta-Mn-O & Hexagonal, 191 & \includegraphics[width=1.5cm]{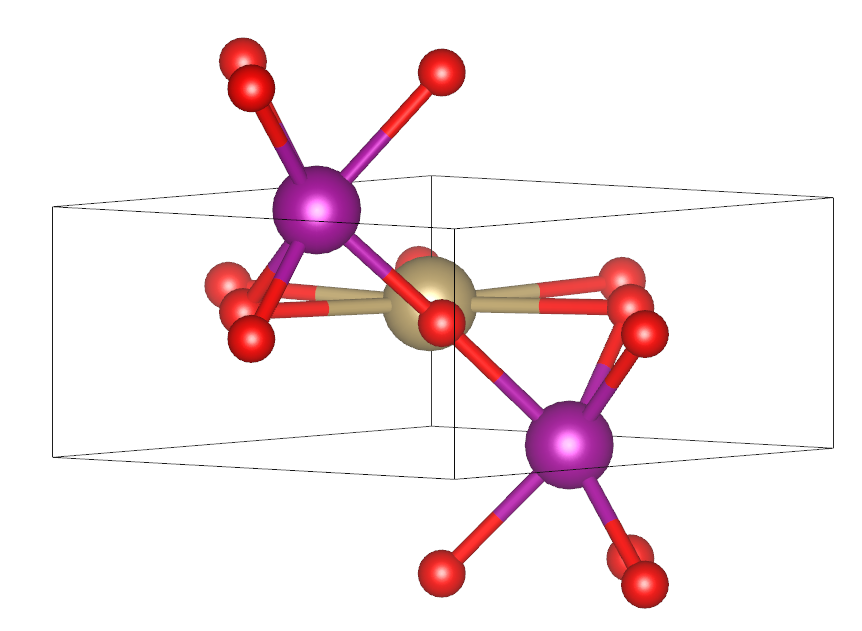}
            & 5.207, 3.12 & 5.1, 3.07 \\
        Cr-Co-P & Orthorhombic, 62 & \includegraphics[width=1.3cm]{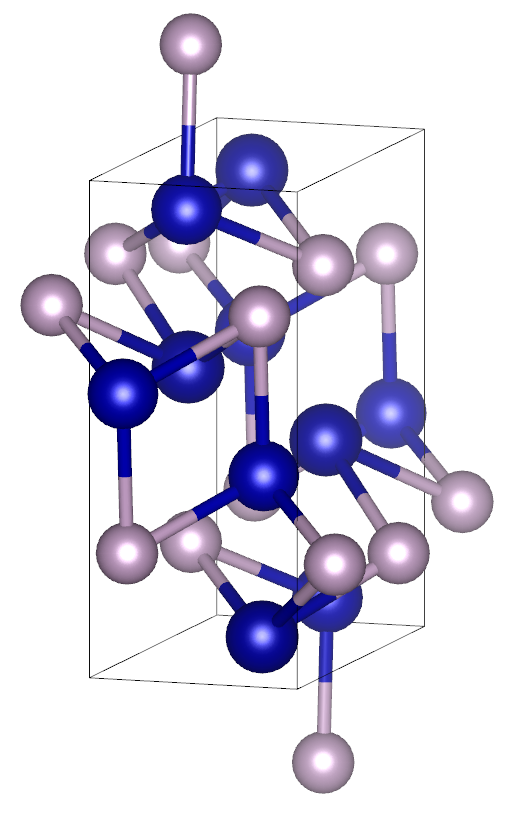}
            & 3.386, 5.889, 6.949 & 3.52, 5.73, 6.68 \\
        Tl-Cd-S & Hexagonal, 187 & \includegraphics[width=1.5cm]{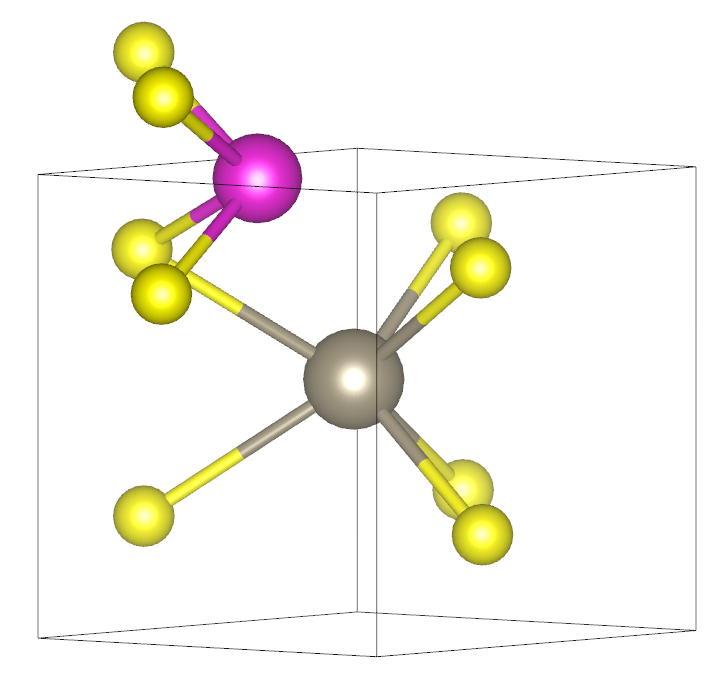}
            & 4.81, 6.33 & 3.89, 7.08 \\
        Ce-In-Cu & Hexagonal, 194 & \includegraphics[width=1.5cm]{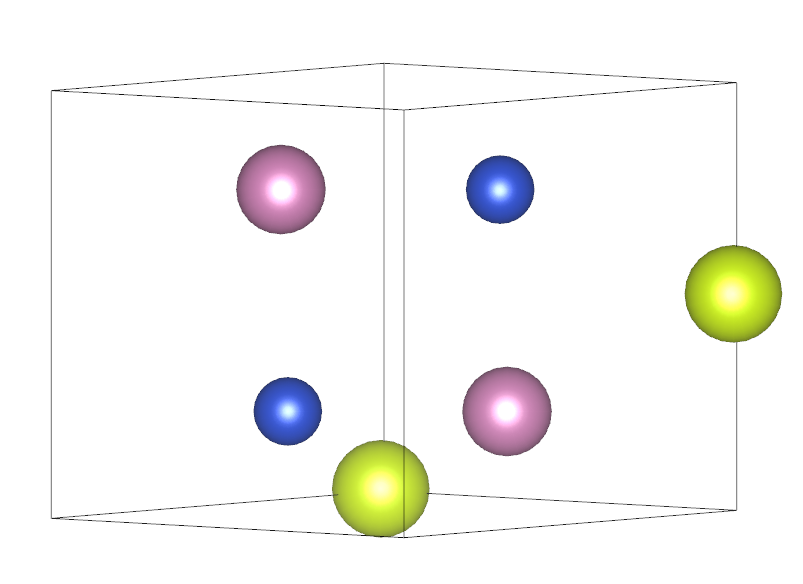}
            & 5.398, 7.592  & 4.74, 6.73 \\
        Mn-Ca-Sn & Tetragonal, 129 & \includegraphics[width=1.5cm]{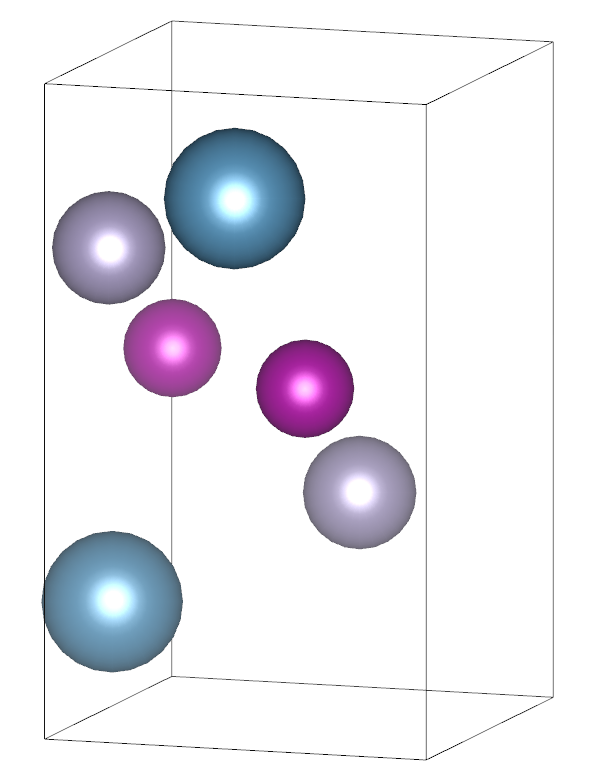}
            & 4.471, 7.387  & 4.53, 7.35 \\
        Ba-Ti-O & Tetragonal, 123 & \includegraphics[width=1.5cm]{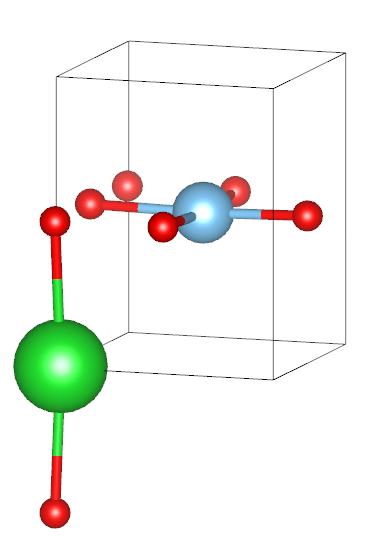}
            & 4.326, 5.587  & 4.11, 5.04 \\
        Pu-I-O & Tetragonal, 129 & \includegraphics[width=1.5cm]{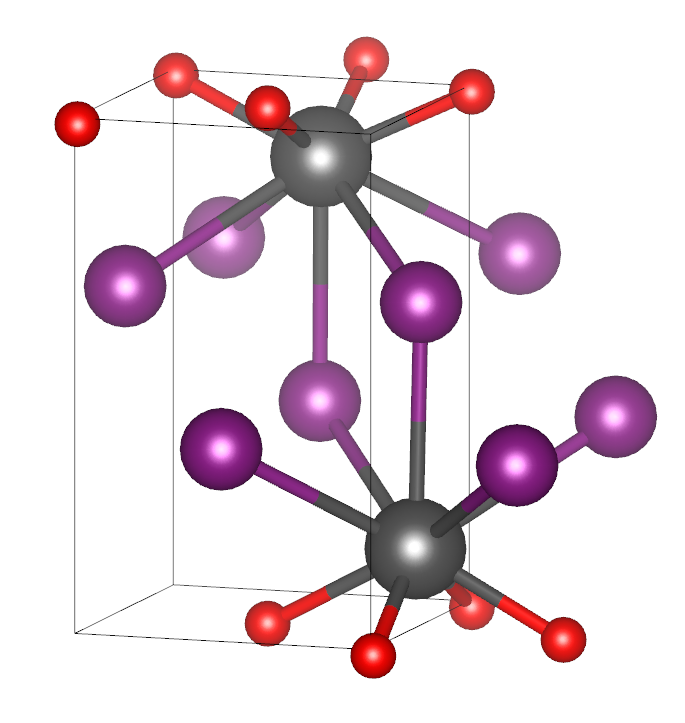}
            & 4.0, 6.7  & 4.01, 9.47 \\
    \end{tblr}
    \caption{Some of the experimentally known structures that were successfully generated by the model when the correct composition and the space group were given as input.}
    \label{tab:reconstruction}
\end{table}

\subsection{Generating new stable structures}


To test the ability to generate new stable crystal structures, we attempt to sample new stable ternary compounds. We chose several chemical systems with no known stable ternary compounds in most of them. As for the space group, we prioritize 32 of the most common (Figure~\ref{space_groups}) space groups, sampling 60 structured per space group per composition. Taking into account the distribution of the number of non-equivalent atomic sites per element (Figure~\ref{mult}), which has to be fixed before sampling, we chose one or two non-equivalent positions for each of the three elements in the composition, with a total of at most four non-equivalent sites per structure.

We have tested our model for several compositions in several systems. Table~\ref{tab:new} displays those that are considered to be the most stable by DFT and we also validated their mechanical stability by additional phonon calculations. 
It turns out that with the estimated formation energy, these structures lie close to the convex hull formed by other known binary configurations composed of the same elements and calculated under the same DFT conditions.

\begin{table}[htbp]
    \begin{tblr}{
      colspec = {X[c,t]X[c]X[c]X[c]X[c]X[c]X[c]},
      stretch = 0,
      hlines = {1.5pt},
      vlines = {1.5pt},
    }
        Chemical system & Crystal family and space group & Generated composition & Heat of Formation, kJ/mol & Energy from the Hull, kJ/mol \\
        Cu-Si-K & Orthorhombic, $Pnma$ & Cu$_2$KSi & -5.29 & +0.61 \\
       Fe-Ta-Sn & Orthorhombic, $Fmmm$ & Fe$_2$TaSn & -8.96 & +4.14 \\        
       Sb-Ta-V & Hexagonal, $P6\bar{3}/mmc$ & SbTa$_2$V$_3$ & -12.55 & +1.57 \\
       Al-K-Pd & Hexagonal, $P6/mmm$ & Al$_3$KPd$_6$ & -55.77 & +15.38 \\            
    \end{tblr}
    \caption{Samples for compositions with no known stable ternary compounds.}
    \label{tab:new}
\end{table}

\begin{figure}
     \centering
     \includegraphics[width=.45\textwidth]{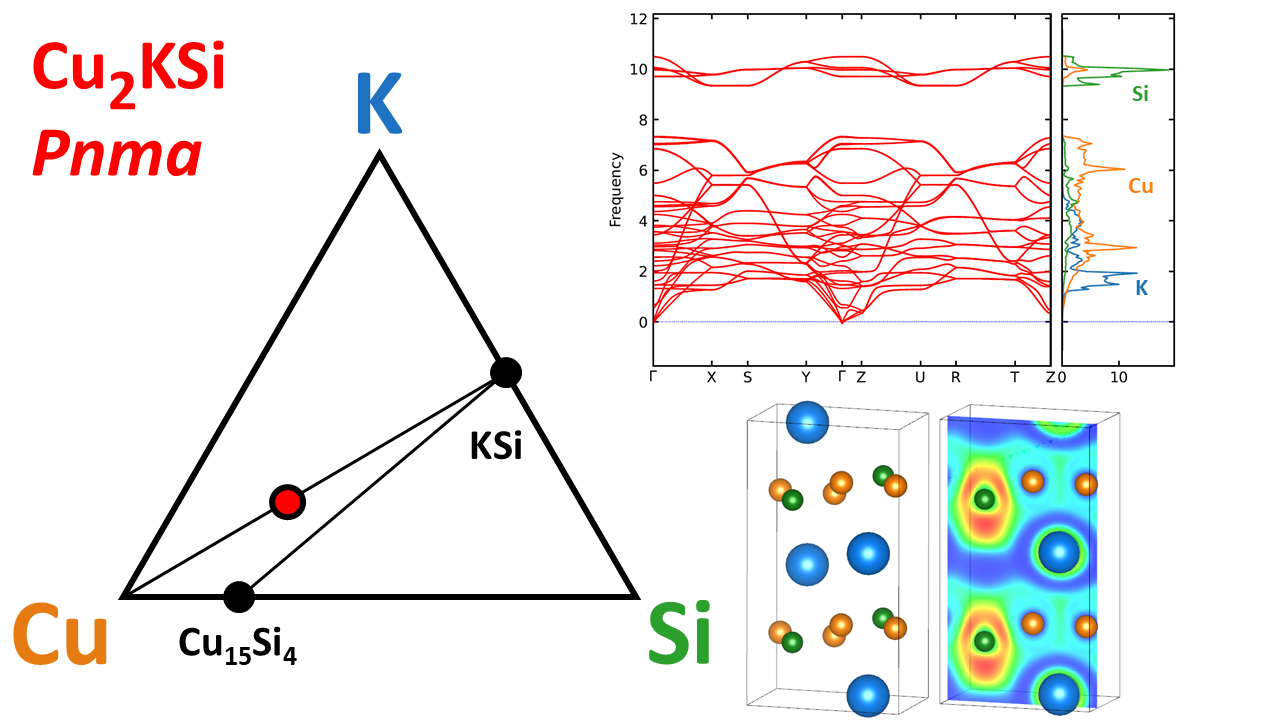}
     \includegraphics[width=.45\textwidth]{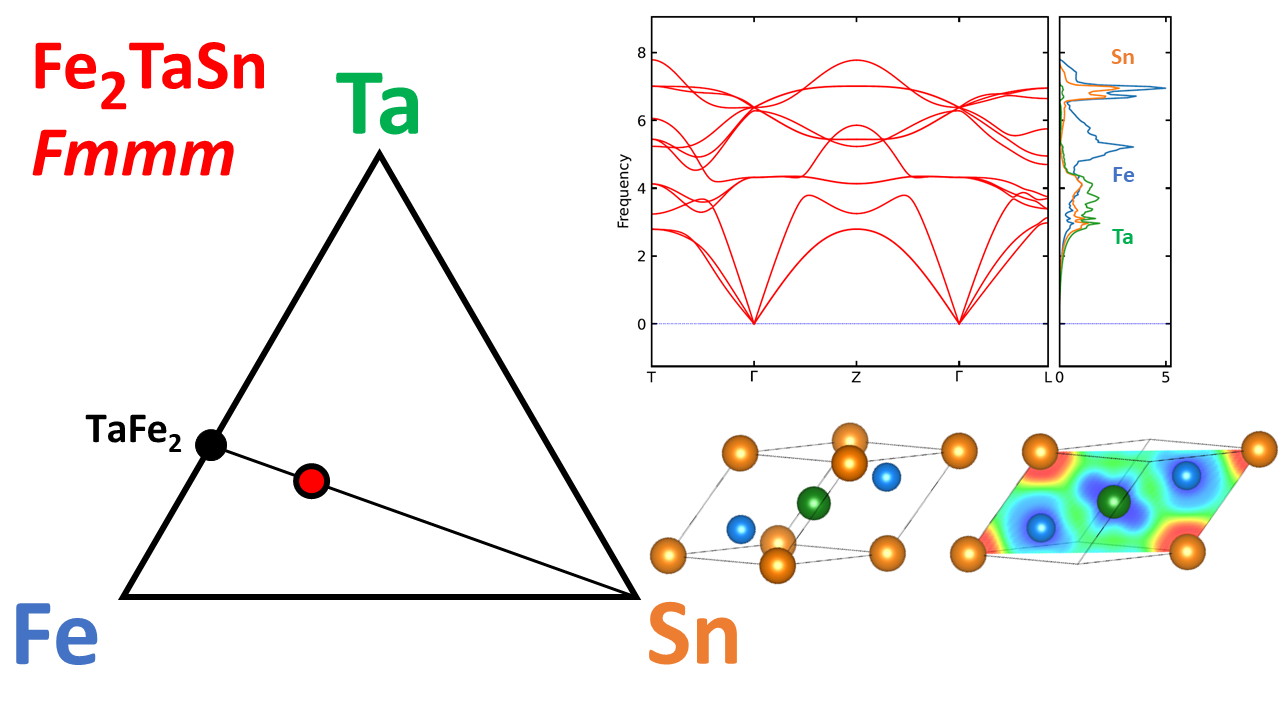}
     \includegraphics[width=.45\textwidth]{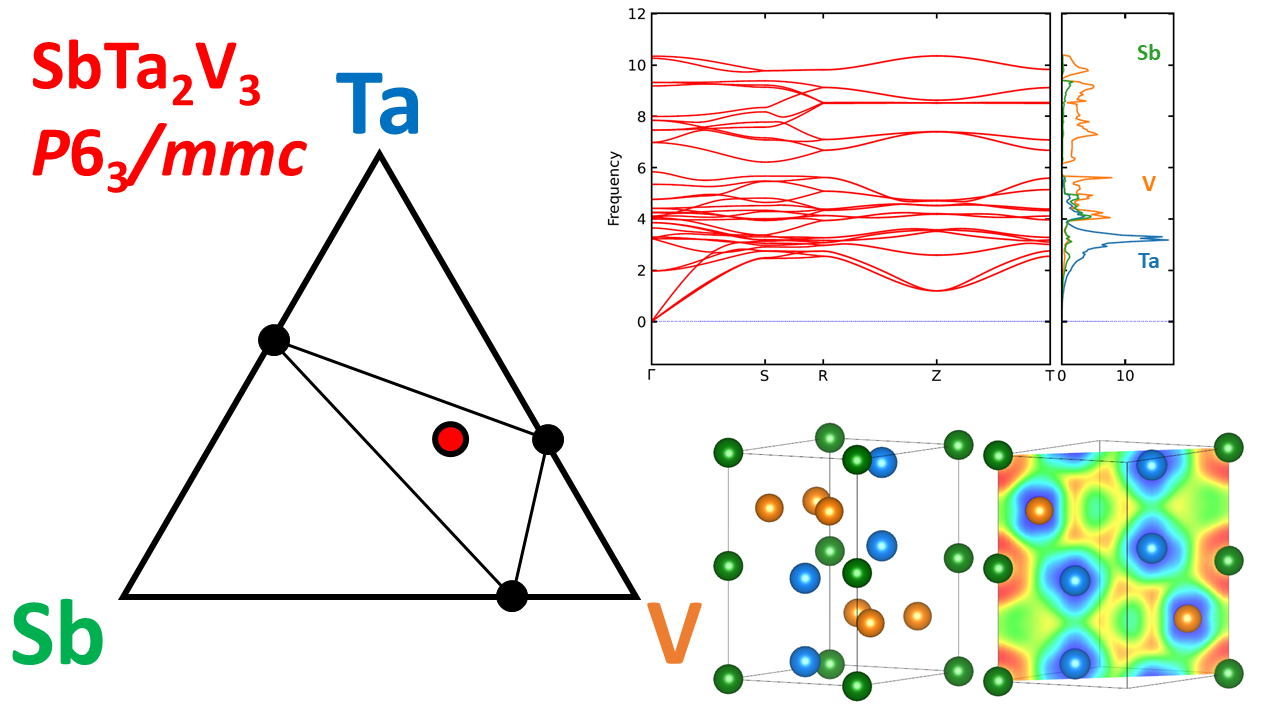}
     \includegraphics[width=.45\textwidth]{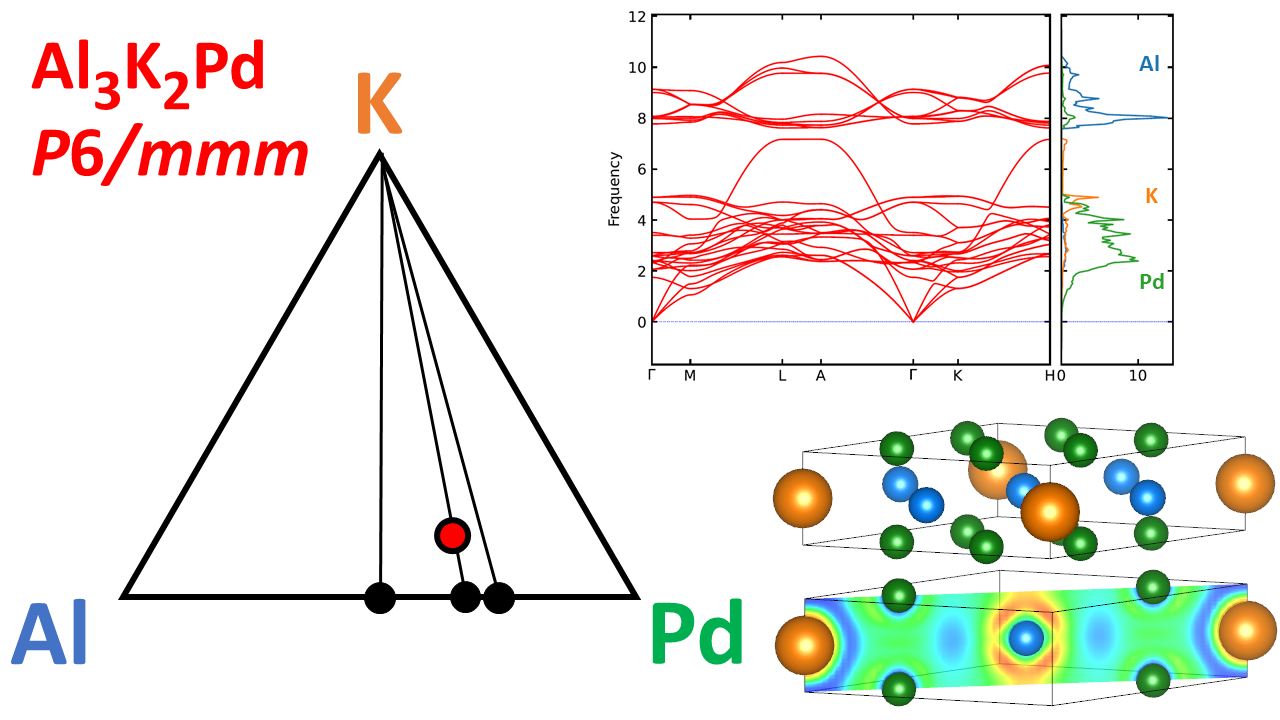}
     \caption{Example of new generated stable compounds verified by first-principles calculations. The phonon dispersion curves indicate no imaginary frequency.
     The electron localization function (ELF) is mapped onto the range 0 (blue) to 1 (red) and represents the probability of finding an electron in the neighbourhood space of a reference electron located at a given point.}
     \label{stable_structures}
\end{figure}

Although no structure seems to support the known ground state, it appears that the generated phases are close to stable and remind us of the difficulty of discovering higher-order compounds (more than binaries, such as ternary compound).
The method demonstrates its great adaptability to a variety of problems by showing that it is able to generate phases with different symmetries.
A closer look at Figure~\ref{stable_structures} also reveals the rich diversity of the proposed lattices, with lamellar structure (Cu$_2$KSi) or more compact phases (SbTa$_2$V$_3$), sometimes magnetic compounds (Fe$_3$Ta$_2$Sn, $M=0.5\,\mu_B$) and a variety of chemical bonds: iono-covalent (Al$_3$KPd$_6$, Cu$_2$KSi), or metallic (Fe$_3$Ta$_2$Sn).


One of the challenges to be addressed in the near future is to speed up the generative process to explore the chemical composition space more extensively. 
However, one has to keep in mind that the verification procedure by DFT calculation remains costly but necessary, as an accurate regression is still not precise enough.

\section{Conclusion and outlook}

Our model successfully applies a score-based generative method to the task of generating new materials. It compares favourably with existing generative models, showing matching or superior results in terms of the studied metrics.

The model is able to reconstruct some of the existing stable structures, in particular it succeeds in denoising the lattice parameters together with the atomic positions. The method performs well across a wide range of space groups and lattice systems, and is not restricted to compositions of a certain number of chemical elements. 

We have also shown that the model is capable of generating entirely new stable structures. 
The model generates structures with a chosen space group and composition, which can be potentially useful in determining the structure of experimentally known materials with unknown unit cells. 
More generally, the model may be combined with any random sampler of compositions and space groups.

It is noteworthy that only the atomic numbers have been entered to indicate the nature of the selected elements, without any additional information such as chemical descriptors that would provide explicit correlations between the properties of different elements. An interesting line of research involves  the integration of such  prior chemical knowledge  to further enhance the model's performance.

At present, sampling a single structure takes several minutes, which in turn makes validation of the model cumbersome. The main bottleneck of the current implementation is the construction of the multigraph, which is performed on CPU and is not yet parallelized.

Another possible valuable extension of the model consists in conditioning the model on the exact composition of the desired structure. At present, the number of non-equivalent sites is fixed in advance, 
but the final ratio of total sites per element still varies widely, which is not always desired.

\section{Data and Software Availability}

The implemented code along with the used data is available online at \url{https://github.com/findooshka/diffusion-atoms}.

\section*{Acknowledgements}
Calculations were performed using HPC resources from GENCI-CINES (No. A0060906175) 
In addition, we acknowledge the financial support from the CNRS-MITI (CRISTAT project).

\bibliography{references}

\providecommand{\latin}[1]{#1}
\makeatletter
\providecommand{\doi}
  {\begingroup\let\do\@makeother\dospecials
  \catcode`\{=1 \catcode`\}=2 \doi@aux}
\providecommand{\doi@aux}[1]{\endgroup\texttt{#1}}
\makeatother
\providecommand*\mcitethebibliography{\thebibliography}
\csname @ifundefined\endcsname{endmcitethebibliography}
  {\let\endmcitethebibliography\endthebibliography}{}
\begin{mcitethebibliography}{53}
\providecommand*\natexlab[1]{#1}
\providecommand*\mciteSetBstSublistMode[1]{}
\providecommand*\mciteSetBstMaxWidthForm[2]{}
\providecommand*\mciteBstWouldAddEndPuncttrue
  {\def\EndOfBibitem{\unskip.}}
\providecommand*\mciteBstWouldAddEndPunctfalse
  {\let\EndOfBibitem\relax}
\providecommand*\mciteSetBstMidEndSepPunct[3]{}
\providecommand*\mciteSetBstSublistLabelBeginEnd[3]{}
\providecommand*\EndOfBibitem{}
\mciteSetBstSublistMode{f}
\mciteSetBstMaxWidthForm{subitem}{(\alph{mcitesubitemcount})}
\mciteSetBstSublistLabelBeginEnd
  {\mcitemaxwidthsubitemform\space}
  {\relax}
  {\relax}

\bibitem[Le and Winkler(2016)Le, and Winkler]{Le16}
Le,~T.; Winkler,~D. Discovery and Optimization of Materials Using Evolutionary
  Approaches. \emph{Chem. Rev.} \textbf{2016}, 6107 -- 6132\relax
\mciteBstWouldAddEndPuncttrue
\mciteSetBstMidEndSepPunct{\mcitedefaultmidpunct}
{\mcitedefaultendpunct}{\mcitedefaultseppunct}\relax
\EndOfBibitem
\bibitem[Henson \latin{et~al.}(2018)Henson, Gromski, and Cronin]{Henson18}
Henson,~A.~B.; Gromski,~P.; Cronin,~L. Designing Algorithms To Aid Discovery by
  Chemical Robots. \emph{ACS Cent. Sci.} \textbf{2018}, 793--804\relax
\mciteBstWouldAddEndPuncttrue
\mciteSetBstMidEndSepPunct{\mcitedefaultmidpunct}
{\mcitedefaultendpunct}{\mcitedefaultseppunct}\relax
\EndOfBibitem
\bibitem[Xie \latin{et~al.}(2020)Xie, Zhang, Hu, Zhang, Kelly, Atwood, and
  Lin]{Xie20}
Xie,~Y.; Zhang,~C.; Hu,~X.; Zhang,~C.; Kelly,~S.~P.; Atwood,~J.~L.; Lin,~J.
  Machine learning assisted synthesis of metal-organic nanocapsules. \emph{J.
  Am. Chem. Soc.} \textbf{2020}, 1475 -- 1481\relax
\mciteBstWouldAddEndPuncttrue
\mciteSetBstMidEndSepPunct{\mcitedefaultmidpunct}
{\mcitedefaultendpunct}{\mcitedefaultseppunct}\relax
\EndOfBibitem
\bibitem[Ward \latin{et~al.}(2016)Ward, Agrawal, Choudhary, and
  Wolverton]{Ward16}
Ward,~L.; Agrawal,~A.; Choudhary,~A.; Wolverton,~C. A general- purpose machine
  learning framework for predicting properties of inorganic materials.
  \emph{npj Comput. Mater.} \textbf{2016}, 16026\relax
\mciteBstWouldAddEndPuncttrue
\mciteSetBstMidEndSepPunct{\mcitedefaultmidpunct}
{\mcitedefaultendpunct}{\mcitedefaultseppunct}\relax
\EndOfBibitem
\bibitem[Dong \latin{et~al.}(2019)Dong, Wu, Zhang, Liu, Cheng, and Lin]{Dong19}
Dong,~Y.; Wu,~C.; Zhang,~C.; Liu,~Y.; Cheng,~J.; Lin,~J. Bandgap prediction by
  deep learning in configurationally hybridized graphene and boron nitride.
  \emph{npj Comput. Mater.} \textbf{2019}, 26\relax
\mciteBstWouldAddEndPuncttrue
\mciteSetBstMidEndSepPunct{\mcitedefaultmidpunct}
{\mcitedefaultendpunct}{\mcitedefaultseppunct}\relax
\EndOfBibitem
\bibitem[Oliynyk \latin{et~al.}(2017)Oliynyk, Adutwum, Rudyk, Pisavadia, Lotfi,
  Hlukhyy, Harynuk, Mar, and Brgoch]{Oliynyk17}
Oliynyk,~A.~O.; Adutwum,~L.~A.; Rudyk,~B.~W.; Pisavadia,~H.; Lotfi,~S.;
  Hlukhyy,~V.; Harynuk,~J.~J.; Mar,~A.; Brgoch,~J. Disentangling Structural
  Confusion through Machine Learning: Structure Prediction and Polymorphism of
  Equiatomic Ternary Phases ABC. \emph{J. Am. Chem. Soc.} \textbf{2017},
  17870--17881\relax
\mciteBstWouldAddEndPuncttrue
\mciteSetBstMidEndSepPunct{\mcitedefaultmidpunct}
{\mcitedefaultendpunct}{\mcitedefaultseppunct}\relax
\EndOfBibitem
\bibitem[Raccuglia \latin{et~al.}(2016)Raccuglia, Elbert, Adler, Falk, Wenny,
  Mollo, Zeller, Friedler, Schrier, and Norquist]{Raccuglia16}
Raccuglia,~P.; Elbert,~K.~C.; Adler,~P.; Falk,~C.; Wenny,~M.; Mollo,~A.;
  Zeller,~M.; Friedler,~S.~A.; Schrier,~J.; Norquist,~A.
  Machine-learning-assisted materials discovery using failed experiments.
  \emph{Nature} \textbf{2016}, 73--76\relax
\mciteBstWouldAddEndPuncttrue
\mciteSetBstMidEndSepPunct{\mcitedefaultmidpunct}
{\mcitedefaultendpunct}{\mcitedefaultseppunct}\relax
\EndOfBibitem
\bibitem[Ahneman \latin{et~al.}(2018)Ahneman, Estrada, Lin, Dreher, and
  Doyle]{Ahneman18}
Ahneman,~D.~T.; Estrada,~J.~G.; Lin,~S.; Dreher,~S.~D.; Doyle,~A. Predicting
  reaction performance in C--N cross-coupling using machine learning.
  \emph{Science} \textbf{2018}, 186--190\relax
\mciteBstWouldAddEndPuncttrue
\mciteSetBstMidEndSepPunct{\mcitedefaultmidpunct}
{\mcitedefaultendpunct}{\mcitedefaultseppunct}\relax
\EndOfBibitem
\bibitem[Granda \latin{et~al.}(2018)Granda, Donina, Dragone, Long, and
  Cronin]{Granda18}
Granda,~J.~M.; Donina,~L.; Dragone,~V.; Long,~D.-L.; Cronin,~L. Controlling an
  organic synthesis robot with machine learning to search for new reactivity.
  \emph{Nature} \textbf{2018}, 377--381\relax
\mciteBstWouldAddEndPuncttrue
\mciteSetBstMidEndSepPunct{\mcitedefaultmidpunct}
{\mcitedefaultendpunct}{\mcitedefaultseppunct}\relax
\EndOfBibitem
\bibitem[Moosavi \latin{et~al.}(2019)Moosavi, Chidambaram, Talirz, Haranczyk,
  Stylianou, and Smit]{Moosavi19}
Moosavi,~S.~M.; Chidambaram,~A.; Talirz,~L.; Haranczyk,~M.; Stylianou,~K.;
  Smit,~B. Capturing chemical intuition in synthesis of metal-organic
  frameworks. \emph{Nat. Commun.} \textbf{2019}, 539\relax
\mciteBstWouldAddEndPuncttrue
\mciteSetBstMidEndSepPunct{\mcitedefaultmidpunct}
{\mcitedefaultendpunct}{\mcitedefaultseppunct}\relax
\EndOfBibitem
\bibitem[Coley \latin{et~al.}(2018)Coley, Green, and Jensen]{Coley18}
Coley,~C.~W.; Green,~W.~H.; Jensen,~K. Machine Learning in Computer-Aided
  Synthesis Planning. \emph{Acc. Chem. Res.} \textbf{2018}, 1281--1289\relax
\mciteBstWouldAddEndPuncttrue
\mciteSetBstMidEndSepPunct{\mcitedefaultmidpunct}
{\mcitedefaultendpunct}{\mcitedefaultseppunct}\relax
\EndOfBibitem
\bibitem[Voznyy \latin{et~al.}(2019)Voznyy, Levina, Fan, Askerka, Jain, Choi,
  Ouellette, Todorovic, Sagar, and Sargent]{Voznyy19}
Voznyy,~O.; Levina,~L.; Fan,~J.~Z.; Askerka,~M.; Jain,~A.; Choi,~M.-J.;
  Ouellette,~O.; Todorovic,~A.~P.; Sagar,~L.; Sargent,~E. Machine Learning
  Accelerates Discovery of Optimal Colloidal Quantum Dot Synthesis. \emph{ACS
  Nano} \textbf{2019}, 11122--11128\relax
\mciteBstWouldAddEndPuncttrue
\mciteSetBstMidEndSepPunct{\mcitedefaultmidpunct}
{\mcitedefaultendpunct}{\mcitedefaultseppunct}\relax
\EndOfBibitem
\bibitem[Kim \latin{et~al.}(2020)Kim, Noh, Go, Aspuru-Guzik, and Jung]{Kim20}
Kim,~S.; Noh,~J.; Go,~G.~H.; Aspuru-Guzik,~A.; Jung,~Y. Generative adversarial
  networks for crystal structure prediction. \emph{ACS Cent. Sci.}
  \textbf{2020}, 1412--1420\relax
\mciteBstWouldAddEndPuncttrue
\mciteSetBstMidEndSepPunct{\mcitedefaultmidpunct}
{\mcitedefaultendpunct}{\mcitedefaultseppunct}\relax
\EndOfBibitem
\bibitem[Noh \latin{et~al.}(2019)Noh, Kim, Stein, Sanchez-Lengeling, Gregoire,
  Aspuru-Guzik, and Jung]{imatgen}
Noh,~J.; Kim,~J.; Stein,~H.~S.; Sanchez-Lengeling,~B.; Gregoire,~J.~M.;
  Aspuru-Guzik,~A.; Jung,~Y. Inverse Design of Solid-State Materials via a
  Continuous Representation. \emph{Matter} \textbf{2019}, \emph{1},
  1370--1384\relax
\mciteBstWouldAddEndPuncttrue
\mciteSetBstMidEndSepPunct{\mcitedefaultmidpunct}
{\mcitedefaultendpunct}{\mcitedefaultseppunct}\relax
\EndOfBibitem
\bibitem[Hoffmann \latin{et~al.}(2019)Hoffmann, Maestrati, Sawada, Tang,
  Sellier, and Bengio]{unet_original}
Hoffmann,~J.; Maestrati,~L.; Sawada,~Y.; Tang,~J.; Sellier,~J.~M.; Bengio,~Y.
  Data-Driven Approach to Encoding and Decoding 3-D Crystal Structures.
  \emph{arXiv preprint} \textbf{2019}, 1909.00949\relax
\mciteBstWouldAddEndPuncttrue
\mciteSetBstMidEndSepPunct{\mcitedefaultmidpunct}
{\mcitedefaultendpunct}{\mcitedefaultseppunct}\relax
\EndOfBibitem
\bibitem[Court \latin{et~al.}(2020)Court, Yildirim, Jain, and Cole]{unet}
Court,~C.~J.; Yildirim,~B.; Jain,~A.; Cole,~J.~M. 3-D Inorganic Crystal
  Structure Generation and Property Prediction via Representation Learning.
  \emph{J. Chem. Inf. Model.} \textbf{2020}, \relax
\mciteBstWouldAddEndPunctfalse
\mciteSetBstMidEndSepPunct{\mcitedefaultmidpunct}
{}{\mcitedefaultseppunct}\relax
\EndOfBibitem
\bibitem[Xie \latin{et~al.}(2022)Xie, Fu, Ganea, Barzilay, and Jaakkola]{CDVAE}
Xie,~T.; Fu,~X.; Ganea,~O.-E.; Barzilay,~R.; Jaakkola,~T.~S. Crystal Diffusion
  Variational Autoencoder for Periodic Material Generation. International
  Conference on Learning Representations. 2022\relax
\mciteBstWouldAddEndPuncttrue
\mciteSetBstMidEndSepPunct{\mcitedefaultmidpunct}
{\mcitedefaultendpunct}{\mcitedefaultseppunct}\relax
\EndOfBibitem
\bibitem[Nouira \latin{et~al.}(2019)Nouira, Sokolovska, and
  Crivello]{crystalgan}
Nouira,~A.; Sokolovska,~N.; Crivello,~J.-C. CrystalGAN: Learning to Discover
  Crystallographic Structures with Generative Adversarial Networks. AAAI 2019
  Spring Symposium on Combining Machine Learning with Knowledge Engineering
  (AAAI-MAKE). 2019\relax
\mciteBstWouldAddEndPuncttrue
\mciteSetBstMidEndSepPunct{\mcitedefaultmidpunct}
{\mcitedefaultendpunct}{\mcitedefaultseppunct}\relax
\EndOfBibitem
\bibitem[Long \latin{et~al.}(2021)Long, Fortunato, Opahle, Zhang, Samathrakis,
  Shen, Gutfleisch, and Zhang]{cgan}
Long,~T.; Fortunato,~N.; Opahle,~I.; Zhang,~Y.; Samathrakis,~I.; Shen,~C.;
  Gutfleisch,~O.; Zhang,~H. Constrained crystals deep convolutional generative
  adversarial network for the inverse design of crystal structures. \emph{npj
  Comput. Mater.} \textbf{2021}, \emph{7}, 66\relax
\mciteBstWouldAddEndPuncttrue
\mciteSetBstMidEndSepPunct{\mcitedefaultmidpunct}
{\mcitedefaultendpunct}{\mcitedefaultseppunct}\relax
\EndOfBibitem
\bibitem[Kim \latin{et~al.}(2020)Kim, Noh, Gu, Aspuru-Guzik, and Jung]{cccgan}
Kim,~S.; Noh,~J.; Gu,~G.~H.; Aspuru-Guzik,~A.; Jung,~Y. Generative Adversarial
  Networks for Crystal Structure Prediction. \emph{ACS Cent. Sci.}
  \textbf{2020}, \relax
\mciteBstWouldAddEndPunctfalse
\mciteSetBstMidEndSepPunct{\mcitedefaultmidpunct}
{}{\mcitedefaultseppunct}\relax
\EndOfBibitem
\bibitem[Han \latin{et~al.}(2023)Han, Lee, Han, Moosavi, Kim, and
  Park]{composition_CDVAE}
Han,~S.; Lee,~J.; Han,~S.; Moosavi,~S.~M.; Kim,~J.; Park,~C. Design of New
  Inorganic Crystals with the Desired Composition Using Deep Learning. \emph{J.
  Chem. Inf. Model.} \textbf{2023}, \emph{63}, 5755--5763, PMID: 37683188\relax
\mciteBstWouldAddEndPuncttrue
\mciteSetBstMidEndSepPunct{\mcitedefaultmidpunct}
{\mcitedefaultendpunct}{\mcitedefaultseppunct}\relax
\EndOfBibitem
\bibitem[Sousa \latin{et~al.}(2021)Sousa, Correia, Pereira, and
  Rocha]{Sousa2021}
Sousa,~T.; Correia,~J.; Pereira,~V.; Rocha,~M. Generative Deep Learning for
  Targeted Compound Design. \emph{J. Chem. Inf. Model.} \textbf{2021},
  \emph{61}, 5343--5361, PMID: 34699719\relax
\mciteBstWouldAddEndPuncttrue
\mciteSetBstMidEndSepPunct{\mcitedefaultmidpunct}
{\mcitedefaultendpunct}{\mcitedefaultseppunct}\relax
\EndOfBibitem
\bibitem[Kang and Cho(2019)Kang, and Cho]{Kang2019}
Kang,~S.; Cho,~K. Conditional Molecular Design with Deep Generative Models.
  \emph{J. Chem. Inf. Model.} \textbf{2019}, \emph{59}, 43--52, PMID:
  30016587\relax
\mciteBstWouldAddEndPuncttrue
\mciteSetBstMidEndSepPunct{\mcitedefaultmidpunct}
{\mcitedefaultendpunct}{\mcitedefaultseppunct}\relax
\EndOfBibitem
\bibitem[Bagal \latin{et~al.}(2022)Bagal, Aggarwal, Vinod, and
  Priyakumar]{MolGPT}
Bagal,~V.; Aggarwal,~R.; Vinod,~P.~K.; Priyakumar,~U.~D. MolGPT: Molecular
  Generation Using a Transformer-Decoder Model. \emph{J. Chem. Inf. Model.}
  \textbf{2022}, \emph{62}, 2064--2076, PMID: 34694798\relax
\mciteBstWouldAddEndPuncttrue
\mciteSetBstMidEndSepPunct{\mcitedefaultmidpunct}
{\mcitedefaultendpunct}{\mcitedefaultseppunct}\relax
\EndOfBibitem
\bibitem[Sattarov \latin{et~al.}(2019)Sattarov, Baskin, Horvath, Marcou,
  Bjerrum, and Varnek]{Sattarov2019}
Sattarov,~B.; Baskin,~I.~I.; Horvath,~D.; Marcou,~G.; Bjerrum,~E.~J.;
  Varnek,~A. De Novo Molecular Design by Combining Deep Autoencoder Recurrent
  Neural Networks with Generative Topographic Mapping. \emph{J. Chem. Inf.
  Model.} \textbf{2019}, \emph{59}, 1182--1196, PMID: 30785751\relax
\mciteBstWouldAddEndPuncttrue
\mciteSetBstMidEndSepPunct{\mcitedefaultmidpunct}
{\mcitedefaultendpunct}{\mcitedefaultseppunct}\relax
\EndOfBibitem
\bibitem[Putin \latin{et~al.}(2018)Putin, Asadulaev, Ivanenkov, Aladinskiy,
  Sanchez-Lengeling, Aspuru-Guzik, and Zhavoronkov]{Putin18}
Putin,~E.; Asadulaev,~A.; Ivanenkov,~Y.; Aladinskiy,~V.; Sanchez-Lengeling,~B.;
  Aspuru-Guzik,~A.; Zhavoronkov,~A. Reinforced adversarial neural computer for
  de novo molecular design. \emph{J. Chem. Inf. Model.} \textbf{2018},
  \emph{58}, 1194\relax
\mciteBstWouldAddEndPuncttrue
\mciteSetBstMidEndSepPunct{\mcitedefaultmidpunct}
{\mcitedefaultendpunct}{\mcitedefaultseppunct}\relax
\EndOfBibitem
\bibitem[Lim \latin{et~al.}(2018)Lim, Ryu, Kim, \latin{et~al.} others]{Lim18}
Lim,~J.; Ryu,~S.; Kim,~J.~W.; others Molecular generative model based on
  conditional variational autoencoder for de novo molecular design. \emph{J.
  Cheminf.} \textbf{2018}, \emph{10}\relax
\mciteBstWouldAddEndPuncttrue
\mciteSetBstMidEndSepPunct{\mcitedefaultmidpunct}
{\mcitedefaultendpunct}{\mcitedefaultseppunct}\relax
\EndOfBibitem
\bibitem[Fuhr and Sumpter(2022)Fuhr, and Sumpter]{Fuhr22}
Fuhr,~A.~S.; Sumpter,~B.~G. Deep generative models for materials discovery and
  machine learning-accelerated innovation. \emph{Front. Mater.} \textbf{2022},
  \emph{8}\relax
\mciteBstWouldAddEndPuncttrue
\mciteSetBstMidEndSepPunct{\mcitedefaultmidpunct}
{\mcitedefaultendpunct}{\mcitedefaultseppunct}\relax
\EndOfBibitem
\bibitem[Pyzer-Knapp \latin{et~al.}(2022)Pyzer-Knapp, Pitera, Staar,
  \latin{et~al.} others]{Pyzer22}
Pyzer-Knapp,~E.~O.; Pitera,~J.~W.; Staar,~P. W.~J.; others Accelerating
  materials discovery using artificial intelligence, high performance computing
  and robotics. \emph{npj Comput. Mater.} \textbf{2022}, \emph{84}\relax
\mciteBstWouldAddEndPuncttrue
\mciteSetBstMidEndSepPunct{\mcitedefaultmidpunct}
{\mcitedefaultendpunct}{\mcitedefaultseppunct}\relax
\EndOfBibitem
\bibitem[Zuo \latin{et~al.}(2021)Zuo, Qin, Chen, Ye, Li, Luo, and Ong]{Zuo21}
Zuo,~Y.; Qin,~M.; Chen,~C.; Ye,~W.; Li,~X.; Luo,~J.; Ong,~S.~P. Accelerating
  materials discovery with Bayesian optimization and graph deep learning.
  \emph{Mater. today} \textbf{2021}, \emph{51}, 126 -- 135\relax
\mciteBstWouldAddEndPuncttrue
\mciteSetBstMidEndSepPunct{\mcitedefaultmidpunct}
{\mcitedefaultendpunct}{\mcitedefaultseppunct}\relax
\EndOfBibitem
\bibitem[Shi \latin{et~al.}(2021)Shi, Luo, Xu, and Tang]{shi}
Shi,~C.; Luo,~S.; Xu,~M.; Tang,~J. Learning Gradient Fields for Molecular
  Conformation Generation. Proceedings of the 38th International Conference on
  Machine Learning. 2021; pp 9558--9568\relax
\mciteBstWouldAddEndPuncttrue
\mciteSetBstMidEndSepPunct{\mcitedefaultmidpunct}
{\mcitedefaultendpunct}{\mcitedefaultseppunct}\relax
\EndOfBibitem
\bibitem[Hoogeboom \latin{et~al.}(2022)Hoogeboom, Satorras, Vignac, and
  Welling]{hoogeboom22}
Hoogeboom,~E.; Satorras,~V.~G.; Vignac,~C.; Welling,~M. Equivariant Diffusion
  for Molecule Generation in 3{D}. Proceedings of the 39th International
  Conference on Machine Learning. 2022; pp 8867--8887\relax
\mciteBstWouldAddEndPuncttrue
\mciteSetBstMidEndSepPunct{\mcitedefaultmidpunct}
{\mcitedefaultendpunct}{\mcitedefaultseppunct}\relax
\EndOfBibitem
\bibitem[Xie and Grossman(2018)Xie, and Grossman]{cgcnn}
Xie,~T.; Grossman,~J.~C. Crystal Graph Convolutional Neural Networks for an
  Accurate and Interpretable Prediction of Material Properties. \emph{Phys.
  Rev. Lett.} \textbf{2018}, \emph{120}\relax
\mciteBstWouldAddEndPuncttrue
\mciteSetBstMidEndSepPunct{\mcitedefaultmidpunct}
{\mcitedefaultendpunct}{\mcitedefaultseppunct}\relax
\EndOfBibitem
\bibitem[Gasteiger \latin{et~al.}(2020)Gasteiger, Giri, Margraf, and
  G{\"u}nnemann]{dimenet++}
Gasteiger,~J.; Giri,~S.; Margraf,~J.~T.; G{\"u}nnemann,~S. Fast and
  Uncertainty-Aware Directional Message Passing for Non-Equilibrium Molecules.
  2020; \url{https://arxiv.org/abs/2011.14115}\relax
\mciteBstWouldAddEndPuncttrue
\mciteSetBstMidEndSepPunct{\mcitedefaultmidpunct}
{\mcitedefaultendpunct}{\mcitedefaultseppunct}\relax
\EndOfBibitem
\bibitem[Gasteiger \latin{et~al.}(2021)Gasteiger, Becker, and
  G\"{u}nnemann]{gemnet}
Gasteiger,~J.; Becker,~F.; G\"{u}nnemann,~S. GemNet: Universal Directional
  Graph Neural Networks for Molecules. Advances in Neural Information
  Processing Systems. 2021; pp 6790--6802\relax
\mciteBstWouldAddEndPuncttrue
\mciteSetBstMidEndSepPunct{\mcitedefaultmidpunct}
{\mcitedefaultendpunct}{\mcitedefaultseppunct}\relax
\EndOfBibitem
\bibitem[Choudhary and DeCost(2021)Choudhary, and DeCost]{alignn}
Choudhary,~K.; DeCost,~B. Atomistic Line Graph Neural Network for improved
  materials property predictions. \emph{npj Comput. Mater.} \textbf{2021},
  \emph{7}\relax
\mciteBstWouldAddEndPuncttrue
\mciteSetBstMidEndSepPunct{\mcitedefaultmidpunct}
{\mcitedefaultendpunct}{\mcitedefaultseppunct}\relax
\EndOfBibitem
\bibitem[Ho \latin{et~al.}(2020)Ho, Jain, and Abbeel]{DDPM}
Ho,~J.; Jain,~A.; Abbeel,~P. Denoising Diffusion Probabilistic Models. Advances
  in Neural Information Processing Systems. 2020; pp 6840--6851\relax
\mciteBstWouldAddEndPuncttrue
\mciteSetBstMidEndSepPunct{\mcitedefaultmidpunct}
{\mcitedefaultendpunct}{\mcitedefaultseppunct}\relax
\EndOfBibitem
\bibitem[Song and Ermon(2019)Song, and Ermon]{song}
Song,~Y.; Ermon,~S. Generative Modeling by Estimating Gradients of the Data
  Distribution. Advances in Neural Information Processing Systems. 2019\relax
\mciteBstWouldAddEndPuncttrue
\mciteSetBstMidEndSepPunct{\mcitedefaultmidpunct}
{\mcitedefaultendpunct}{\mcitedefaultseppunct}\relax
\EndOfBibitem
\bibitem[Yang \latin{et~al.}(2022)Yang, Zhang, Hong, Xu, Zhao, Shao, Zhang,
  Yang, and Cui]{diffusion_survey}
Yang,~L.; Zhang,~Z.; Hong,~S.; Xu,~R.; Zhao,~Y.; Shao,~Y.; Zhang,~W.;
  Yang,~M.-H.; Cui,~B. Diffusion Models: A Comprehensive Survey of Methods and
  Applications. 2022; \url{https://arxiv.org/abs/2209.00796}\relax
\mciteBstWouldAddEndPuncttrue
\mciteSetBstMidEndSepPunct{\mcitedefaultmidpunct}
{\mcitedefaultendpunct}{\mcitedefaultseppunct}\relax
\EndOfBibitem
\bibitem[Xu \latin{et~al.}(2022)Xu, Yu, Song, Shi, Ermon, and Tang]{Xu22}
Xu,~M.; Yu,~L.; Song,~Y.; Shi,~C.; Ermon,~S.; Tang,~J. Geo{D}iff: a Geometric
  Diffusion Model for Molecular Conformation Generation. International
  Conference on Learning Representations. 2022\relax
\mciteBstWouldAddEndPuncttrue
\mciteSetBstMidEndSepPunct{\mcitedefaultmidpunct}
{\mcitedefaultendpunct}{\mcitedefaultseppunct}\relax
\EndOfBibitem
\bibitem[Jain \latin{et~al.}(2013)Jain, Ong, Hautier, Chen, Richards, Dacek,
  Cholia, Gunter, Skinner, Ceder, and Persson]{MP}
Jain,~A.; Ong,~S.~P.; Hautier,~G.; Chen,~W.; Richards,~W.~D.; Dacek,~S.;
  Cholia,~S.; Gunter,~D.; Skinner,~D.; Ceder,~G.; Persson,~K.~A. Commentary:
  The Materials Project: A materials genome approach to accelerating materials
  innovation. \emph{APL Mater.} \textbf{2013}, \emph{1}, 011002\relax
\mciteBstWouldAddEndPuncttrue
\mciteSetBstMidEndSepPunct{\mcitedefaultmidpunct}
{\mcitedefaultendpunct}{\mcitedefaultseppunct}\relax
\EndOfBibitem
\bibitem[Ren \latin{et~al.}(2020)Ren, Noh, Tian, Oviedo, Xing, Liang, Aberle,
  Liu, Li, Jayavelu, \latin{et~al.} others]{Ren20}
Ren,~Z.; Noh,~J.; Tian,~S.; Oviedo,~F.; Xing,~G.; Liang,~Q.; Aberle,~A.;
  Liu,~Y.; Li,~Q.; Jayavelu,~S.; others Inverse design of crystals using
  generalized invertible crystallographic representation. \emph{arXiv preprint
  arXiv:2005.07609} \textbf{2020}, \relax
\mciteBstWouldAddEndPunctfalse
\mciteSetBstMidEndSepPunct{\mcitedefaultmidpunct}
{}{\mcitedefaultseppunct}\relax
\EndOfBibitem
\bibitem[Baird \latin{et~al.}(2022)Baird, Jablonka, Alverson, Sayeed, Khan,
  Seegmiller, Smit, and Sparks]{Baird22}
Baird,~S.~G.; Jablonka,~K.~M.; Alverson,~M.~D.; Sayeed,~H.~M.; Khan,~M.~F.;
  Seegmiller,~C.; Smit,~B.; Sparks,~T.~D. xtal2png: A python package for
  representing crystal structure as png files. \emph{J. Open Source Softw.}
  \textbf{2022}, \emph{7}, 4528\relax
\mciteBstWouldAddEndPuncttrue
\mciteSetBstMidEndSepPunct{\mcitedefaultmidpunct}
{\mcitedefaultendpunct}{\mcitedefaultseppunct}\relax
\EndOfBibitem
\bibitem[Zhao \latin{et~al.}(2021)Zhao, Al-Fahdi, Hu, Siriwardane, Song,
  Nasiri, and Hu]{Zhao21}
Zhao,~Y.; Al-Fahdi,~M.; Hu,~M.; Siriwardane,~E.; Song,~Y.; Nasiri,~A.; Hu,~J.
  High-throughput discovery of novel cubic crystal materials using deep
  generative neural networks. \emph{Adv. Sci.} \textbf{2021}, \emph{8}\relax
\mciteBstWouldAddEndPuncttrue
\mciteSetBstMidEndSepPunct{\mcitedefaultmidpunct}
{\mcitedefaultendpunct}{\mcitedefaultseppunct}\relax
\EndOfBibitem
\bibitem[Siriwardane \latin{et~al.}(2022)Siriwardane, Zhao, Perera, and
  Hu]{Siriwardane22}
Siriwardane,~E.; Zhao,~Y.; Perera,~I.; Hu,~J. Generative design of stable
  semiconductor materials using deep learning and dft. \emph{npj Comput.
  Mater.} \textbf{2022}, \emph{8}, 164\relax
\mciteBstWouldAddEndPuncttrue
\mciteSetBstMidEndSepPunct{\mcitedefaultmidpunct}
{\mcitedefaultendpunct}{\mcitedefaultseppunct}\relax
\EndOfBibitem
\bibitem[Aroyo \latin{et~al.}(2006)Aroyo, Perez-Mato, Capillas, Kroumova,
  Ivantchev, Madariaga, Kirov, and Wondratschek]{bilbao}
Aroyo,~M.~I.; Perez-Mato,~J.~M.; Capillas,~C.; Kroumova,~E.; Ivantchev,~S.;
  Madariaga,~G.; Kirov,~A.; Wondratschek,~H. Bilbao Crystallographic Server: I.
  Databases and crystallographic computing programs. \emph{Z. Kristallogr. -
  Cryst. Mater.} \textbf{2006}, \emph{221}, 15--27\relax
\mciteBstWouldAddEndPuncttrue
\mciteSetBstMidEndSepPunct{\mcitedefaultmidpunct}
{\mcitedefaultendpunct}{\mcitedefaultseppunct}\relax
\EndOfBibitem
\bibitem[Vincent(2011)]{score_proof}
Vincent,~P. A Connection Between Score Matching and Denoising Autoencoders.
  \emph{Neural Comput.} \textbf{2011}, \emph{23}, 1661--1674\relax
\mciteBstWouldAddEndPuncttrue
\mciteSetBstMidEndSepPunct{\mcitedefaultmidpunct}
{\mcitedefaultendpunct}{\mcitedefaultseppunct}\relax
\EndOfBibitem
\bibitem[Togo and Tanaka(2015)Togo, and Tanaka]{phonopy}
Togo,~A.; Tanaka,~I. First principles phonon calculations in materials science.
  \emph{Scr. Mater.} \textbf{2015}, \emph{108}, 1--5\relax
\mciteBstWouldAddEndPuncttrue
\mciteSetBstMidEndSepPunct{\mcitedefaultmidpunct}
{\mcitedefaultendpunct}{\mcitedefaultseppunct}\relax
\EndOfBibitem
\bibitem[Zhao \latin{et~al.}(2023)Zhao, Siriwardane, Wu, Fu, Al-Fahdi, Hu, and
  Hu]{PGCGM}
Zhao,~Y.; Siriwardane,~E. M.~D.; Wu,~Z.; Fu,~N.; Al-Fahdi,~M.; Hu,~M.; Hu,~J.
  Physics guided deep learning for generative design of crystal materials with
  symmetry constraints. \emph{npj Comput. Mater.} \textbf{2023}, 73--76\relax
\mciteBstWouldAddEndPuncttrue
\mciteSetBstMidEndSepPunct{\mcitedefaultmidpunct}
{\mcitedefaultendpunct}{\mcitedefaultseppunct}\relax
\EndOfBibitem
\bibitem[Belsky \latin{et~al.}(2002)Belsky, Hellenbrandt, Karen, and
  Luksch]{ICSD}
Belsky,~A.; Hellenbrandt,~M.; Karen,~V.; Luksch,~P. New Developments in the
  Inorganic Crystal Structure Database (ICSD): Accessibility in Support of
  Materials Research and Design. \emph{Acta Crystallogr., Sect. B: Struct.
  Sci.} \textbf{2002}, \emph{58}, 364--9\relax
\mciteBstWouldAddEndPuncttrue
\mciteSetBstMidEndSepPunct{\mcitedefaultmidpunct}
{\mcitedefaultendpunct}{\mcitedefaultseppunct}\relax
\EndOfBibitem
\bibitem[Kirklin \latin{et~al.}(2015)Kirklin, Saal, Meredig, Thompson, Doak,
  Aykol, R{\"u}hl, and Wolverton]{OQMD}
Kirklin,~S.; Saal,~J.~E.; Meredig,~B.; Thompson,~A.; Doak,~J.~W.; Aykol,~M.;
  R{\"u}hl,~S.; Wolverton,~C. The Open Quantum Materials Database (OQMD):
  assessing the accuracy of DFT formation energies. \emph{npj Comput. Mater.}
  \textbf{2015}, \emph{1}, 15010\relax
\mciteBstWouldAddEndPuncttrue
\mciteSetBstMidEndSepPunct{\mcitedefaultmidpunct}
{\mcitedefaultendpunct}{\mcitedefaultseppunct}\relax
\EndOfBibitem
\bibitem[Heusel \latin{et~al.}(2018)Heusel, Ramsauer, Unterthiner, Nessler, and
  Hochreiter]{TTUR}
Heusel,~M.; Ramsauer,~H.; Unterthiner,~T.; Nessler,~B.; Hochreiter,~S. GANs
  Trained by a Two Time-Scale Update Rule Converge to a Local Nash Equilibrium.
  2018; \url{https://arxiv.org/abs/1706.08500}\relax
\mciteBstWouldAddEndPuncttrue
\mciteSetBstMidEndSepPunct{\mcitedefaultmidpunct}
{\mcitedefaultendpunct}{\mcitedefaultseppunct}\relax
\EndOfBibitem
\end{mcitethebibliography}

\end{document}